\documentclass[12pt]{article}
\usepackage{amssymb,amsmath,amscd}
\makeatletter

\@addtoreset{equation}{section}
\def\section{\@startsection {section}{1}{\z@}{-2.25ex plus -1ex minus
 -.2ex}{1.0ex plus .2ex}{\large\bf}}
\def\subsection{\@startsection{subsection}{2}{\z@}{-2.0ex plus%
 -1ex minus -.2ex}{0.5ex plus .2ex}{\bf}}

\def\Rea{\mbox{Re}}
\def\Ima{\mbox{Im}}
\def\Iso{\mbox{I}}
\def\Ad{\mathrm{Ad}}

\newcommand{\inv}[0]{{-1}}

\newcommand{\cif}[0]{\mathcal{C}^\infty}

\newcommand{\cd}[0]{\!\cdot\!}

\newcommand{\oo}[0]{\otimes}

\newcommand{\norp}[0]{N_+}
\newcommand{\norm}[0]{N_-}
\newcommand{\Sq}{S}

\def\bl{{\mbox{\boldmath $l$}}}
\def\bv{{\mbox{\boldmath $v$}}}
\def\bw{{\mbox{\boldmath $w$}}}

\def\bx{{\mbox{\boldmath $x$}}}

\def\by{{\mbox{\boldmath $y$}}}

\def\bk{{\mbox{\boldmath $k$}}}
\def\bp{{\mbox{\boldmath $p$}}}

\def\bq{{\mbox{\boldmath $q$}}}
\def\be{{\mbox{\boldmath $e$}}}

\def\bJ{{\mbox{\boldmath $J$}}}
\def\bS{{\mbox{\boldmath $S$}}}

\def\bn{{\mbox{\boldmath $n$}}}
\def\tbn{\tilde \bn}
\def\bm{{\mbox{\boldmath $m$}}}

\newcommand{\gothg}{\mathfrak g }

\newcommand{\gothh}{\mathfrak h }

\newcommand{\an}{\mathfrak{an}}

\newcommand{\RR}{\mathbb{R}}
\newcommand{\CC}{\mathbb{C}}

\newcommand{\HH}{\mathbb{H}}

\newtheorem{theorem}{Theorem}[section]
\newtheorem{lemma}[theorem]{Lemma}
\newtheorem{corollary}[theorem]{Corollary}

\newtheorem{definition}[theorem]{Definition}

\def\bea{\begin{eqnarray}}
\def\eea{\end{eqnarray}}
\def\bmz{\left(\begin{array}{2,2}}
\def\emz{\end{array}\right)}
\def\bmd{\left(\begin{array}{3,3}}
\def\emd{\end{array}\right)}

\newcommand{\tl}[0]{\theta}
\def\bpm{\begin{pmatrix}}
\def\epm{\end{pmatrix}}

\begin{document}
\parskip 6pt
\parindent 0pt
\begin{flushright}
EMPG-07-15\\
pi-qg-55\\
\end{flushright}

\begin{center}
\baselineskip 24 pt {\Large \bf  Quaternionic and Poisson-Lie
structures in 3d gravity: the cosmological constant as deformation
parameter}

\baselineskip 16 pt

\vspace{.7cm} {{ C.~Meusburger}\footnote{\tt  cmeusburger@perimeterinstitute.ca}\\
Perimeter Institute for Theoretical Physics\\
31 Caroline Street North,
Waterloo, Ontario N2L 2Y5, Canada\\

\vspace{.5cm}
{ B.~J.~Schroers}\footnote{\tt bernd@ma.hw.ac.uk} \\
Department of Mathematics and Maxwell Institute for Mathematical Sciences \\
 Heriot-Watt University \\
Edinburgh EH14 4AS, United Kingdom } \\

\vspace{0.5cm}

{August   2007}

\end{center}

\begin{abstract}
\noindent  Each of the  local isometry groups arising in
3d gravity can be viewed as the group of unit
(split) quaternions over a ring  which depends on the cosmological constant.
In this paper we explain and prove this statement, and use it as
a unifying framework for studying Poisson structures associated
with the local isometry groups. We show that,
in all cases except for Euclidean signature with positive cosmological
constant, the local isometry groups are equipped with the Poisson-Lie structure
of a classical double. We calculate the dressing action of the factor
groups on each other and find, amongst others, a simple and unified
description of the symplectic leaves of $SU(2)$ and $SL(2,\RR)$.
We also compute the Poisson structure
on the dual Poisson-Lie groups of the local isometry groups  and on their
Heisenberg doubles; together, they determine the Poisson  structure
of the phase space of 3d gravity in the so-called combinatorial description.
\end{abstract}

\centerline{PACS numbers: 04.20.Cv, 02.20.Qs, 02.40.-k}

\section{Introduction}
In  3d gravity, solutions of the Einstein equations are locally
isometric to a model spacetime which is determined by the signature
of spacetime (Euclidean or Lorentzian) and the cosmological constant
\cite{Carlipbook}. The isometry groups of these  model spacetimes
are therefore
 local isometry groups in 3d gravity. In the formulation of 3d gravity as a Chern-Simons
gauge theory \cite{AT,Witten1}, the local isometry groups play the role of gauge groups.
For Euclidean gravity, the relevant  groups
are $SU(2)\times SU(2)$ for positive cosmological constant,
$SL(2,\CC)$ for negative cosmological constant and the (double cover of the)
Euclidean group $SU(2)\ltimes \RR^3$ for vanishing cosmological constant.
With Lorentzian signature the relevant local isometry groups are
$SL(2,\RR)\times SL(2,\RR)$ for negative cosmological constant,
$SL(2,\CC)$ for positive cosmological constant and the (double cover of the)
Poincar\'e group $SL(2,\RR)\ltimes \RR^3$ for vanishing cosmological constant.
These groups are structurally quite diverse:
some  are direct products of real, simple Lie groups, some are
 semi-direct products and one is a complex Lie group.  As a result,  the techniques used
in the literature on 3d gravity differ widely, with different
approaches  taken for different signatures  and values of the
cosmological constant, which makes it difficult to relate these
different cases and to establish a unified and coherent picture of
the theory.

This paper is motivated by the desire for  one framework
encompassing  all the different signatures and values of the
cosmological constant. The unified description of the isometry
groups and their Lie  algebras in \cite{ich3} in terms of a ring
which depends on the cosmological constant suggests that this may be
possible. Here we begin by  describing the local isometry  groups in
terms of quaternions and  the ring introduced in \cite{ich3}. The
quaternionic description of the local isometry groups generalises
the well-known fact that $SU(2)$ is isomorphic to the unit
quaternions and that $SL(2,\CC)$  is the complexification of
$SU(2)$, i.e.  isomorphic to the unit quaternions with complex
coefficients. For the Lorentzian setting one uses a Lorentzian
version of the quaternions,  called split quaternions, and
the fact that the group of unit split quaternions is isomorphic to
$SL(2,\RR)$; it is then easy to see that $SL(2,\CC)$ can also be
viewed as the set of unit split quaternions with complex
coefficients. To obtain the remaining groups one needs to generalise
$\CC$ to a ring which, depending on the sign of the cosmological
constant and the signature, is isomorphic to $\CC$, the so-called
dual numbers or the split complex numbers.  The upshot of this
construction is a unified description of the local isometry groups
with the cosmological constant appearing as a deformation parameter.

The formulation of 3d gravity not only requires a choice of the
local isometry group but also of an invariant inner product on the
Lie algebra of this group.
This is most readily apparent in the Chern-Simons formulation of the
theory,   where triad and spin-connection are combined into a
Chern-Simons gauge field and the inner product enters the action
explicitly. The inner product
 ultimately determines the symplectic
structure of the phase space. For universes of topology $\RR\times
S$, where $S$ is a surface of arbitrary genus and possibly with a number
of punctures, the phase space can be studied very effectively in the
Hamiltonian or combinatorial approach \cite{FR,AMII,AS}.
In this approach, the phase space is realised as a
quotient of an auxiliary, finite dimensional space. The auxiliary
space has a Poisson structure which is determined by a classical
$r$-matrix whose symmetric part equals the inner product used in
the definition of the Chern-Simons action. We will not review
the  details of this
construction and its application to 3d gravity  \cite{schroers,BNR,we1,we2} here,
but to motivate the second half of this paper we note that
the classical $r$-matrix can be used to define a variety of Poisson
structures, three of which play a fundamental role in the combinatorial approach.

First of all there is  the so-called Sklyanin bracket \cite{CP} on
the isometry group,  endowing it  with the structure of a
Poisson-Lie group; in this paper we will describe this structure
 in the unified, quaternionic language introduced above
for all the isometry groups arising in 3d gravity with the exception of
$SU(2)\times SU(2)$, where the required $r$-matrix does not exist.
 Every Poisson-Lie group has a dual Poisson-Lie group, where
Lie brackets and Poisson brackets are, in a suitable sense,
interchanged. The Poisson-Lie groups
 arising in 3d gravity have the special property that
 in each case the  dual Poisson-Lie group is diffeomorphic
to the original one. As a result, one can  define
 a second Poisson bracket, sometimes  called the dual or Semenov-Tian-Shansky
bracket \cite{setishan}, on the original group.  The symplectic
leaves of the dual Poisson structure are the conjugacy classes of
the original group. In the combinatorial description of the phase
space of Chern-Simons theory this second Poisson structure is
associated with the punctures on the surface $S$, which represent
gravitationally interacting massive point particles with spin.
Finally, the third Poisson structure which arises in the
combinatorial description is the called the Heisenberg double
structure \cite{setishan}; it is a symplectic Poisson structure
defined on two copies of the original group and  is associated with handles of the
surface $S$ \cite{AMII}.

While our work is motivated by its potential use in 3d gravity,
it also has interesting ramifications in pure mathematics.
One is related to the fact that all Poisson-Lie groups discussed
in this paper are classical doubles. Elements in a neighbourhood
of  the identity (and in some cases in the entire group)
can be written uniquely as a product of two elements belonging
to  a pair of Poisson-Lie subgroups. One of the subgroups  is isomorphic to
$SU(2)$ in the Euclidean cases and isomorphic to $SL(2,\RR)$ in
 the Lorentzian cases; the other subgroup
is either $\RR^3$ or  isomorphic to the
group of $2\times 2$  matrices of the form
\bea
\bpm e^{\alpha} & x+iy\\0 & e^{-\alpha}\epm, \qquad \alpha,x,y\in \RR.
\eea
The group of such matrices will be denoted by $AN(2)$ in this paper.
The cases where one factor is $\RR^3$ are degenerate and  rather trivial
limits of the generic situation where one of the factor groups
is isomorphic to $AN(2)$. We therefore consider the generic situation in this paper,
and study the mutual dressing actions of the Poisson-Lie groups
$AN(2)$ on the one hand  and $SU(2)$ or, respectively, $SL(2,\RR)$
on the other. The dressing
actions are defined by comparing the factorisation in one order
with the factorisation in the other order, and the orbits under the
dressing actions give the symplectic leaves of the Poisson structures.
Our results relate the mutual dressing actions of $SU(2)$
and $AN(2)$, discussed in many textbooks  \cite{CP,Majid}
to the mutual dressing actions of $SL(2,\RR)$ and $AN(2)$.  In particular,
our treatment leads to a unified description of the symplectic
leaves of $SU(2)$ and $SL(2,\RR)$ for the various choices of Poisson-Lie structures.

The plan of the paper is as follows. In Sect.~2 we introduce basic
notation and review the result of \cite{ich3} that
 the Lie algebras of the local isometry groups arising
in 3d gravity can be obtained by tensoring $\mathfrak{su}(2)$
or $\mathfrak{sl}(2)$ with a ring $R_\Lambda$ which depends on the cosmological
constant. In Sect.~3 we give a unified description of the local
isometry groups in terms of (split) quaternions and the ring $R_\Lambda$;
we introduce an involution on these  groups and show
that the fixed point sets of this involution are isomorphic to the
model spacetimes arising in 3d gravity.
 Sect.~4 contains a description of double structure of
the isometry groups  and a fundamental theorem 
about the factorisation of a general element
into elements of the two  subgroups described above.
In Sect.~5 we use the factorisation theorem to
define dressing actions  of the subgroups  on each other,
 and we  study the geometry of dressing orbits.  The final Sect.~6
contains a unified description of the Sklyanin, dual and Heisenberg double
Poisson structures associated to the Poisson-Lie groups arising in 3d gravity.
It also contains the characterisation of the dressing orbits studied in Sect.~5
as symplectic leaves of Poisson-Lie group structures on
$AN(2)$ and  $SU(2)$ respectively $SL(2,\RR)$.

\section{The Lie algebras of 3d gravity}

\subsection{Notation and conventions} Throughout the paper we use
Einstein's summation convention. Indices are raised and lowered with
either the three-dimensional Euclidean metric
$\eta^E=\text{diag}(1,1,1)$ or the three-dimensional Minkowski
metric $\eta^L=\text{diag}(1,-1,-1)$. Where necessary, we specify
the signature by a superscript $E$ for Euclidean and $L$ for Lorentzian
signature, which we omit in formulas valid for both signatures. In
particular, we write
\begin{align}
\bp\cd\bq=\eta_{ab}p^aq^b,\qquad\mbox{with}\quad \bp=(p^0,p^1,p^2),
\bq=(q^0,q^1,q^2)\in\RR^3,
\end{align}
where $\eta$ is either the three-dimensional Euclidean or the
three-dimensional Minkowski metric. We also sometimes write simply $ \bp\bq$ for $\bp\cd\bq$
 and $\bp^2$ for $\bp\cd \bp$.

We denote by $J^E_a$, $a=0,1,2$, and  $J^L_a$,
$a=0,1,2$, respectively, the generators of the three-dimensional
rotation algebra $\mathfrak{su}(2)$ and the three-dimensional
Lorentz algebra $\mathfrak{sl}(2,\RR)$. In terms of these generators
the Lie bracket and Killing form are
\begin{align}
\label{lorbr} [J_a,J_b]=\epsilon_{abc}J^c,\qquad
\kappa(J_a,J_b)=\eta_{ab},
\end{align}
where indices are raised and lowered with the metrics
$\eta=\eta^E$ or  $\eta=\eta^L$, and $\epsilon$ denotes the fully
antisymmetric tensor in three dimensions with the convention
$\epsilon_{012}=\epsilon^{012}=1$ (for both signatures).

\subsection{ Lie algebras over a ring}

In this subsection, we assemble some well-known facts and
definitions for the Lie algebras occurring in 3d
gravity as well as some more recent results from \cite{ich3}. As
shown by Witten \cite{Witten1} the Lie algebras arising in
3d  gravity can be expressed in a common form in which
the cosmological constant $\Lambda_c$  appears as a parameter in the
Lie bracket. Defining
\bea \label{Lambdadef} \Lambda = \left\{
\begin{array}{l l}
 \Lambda_c  & \mbox{for Euclidean signature} \\
   -\Lambda_c &\mbox{for Lorentzian signature}
\end{array}
\right.
\eea
 these Lie algebras, in the following denoted by
$\gothh_\Lambda$, are the six-dimensional Lie algebras with
generators $J_a,P_a$, $a=0,1,2$, and Lie brackets\footnote{Our parameter $\Lambda$ is called $\lambda$ in
\cite{Witten1}.}
\begin{align}
\label{liebra} [J_a,J_b]=\epsilon_{abc} J^c,\qquad
[J_a,P_b]=\epsilon_{abc} P^c,\qquad
[P_a,P_b]=\Lambda\epsilon_{abc}J^c.
\end{align}
Again, indices are raised and lowered with the three-dimensional
Euclidean metric $\eta=\eta^E$  or with the Minkowski metric
$\eta=\eta^L$. For  $\Lambda=0$, the bracket of the generators $P_a$
vanishes, and the Lie algebra $\gothh_\Lambda$ is the
three-dimensional Euclidean and Poincar\'e algebra. For $\Lambda<0$,
one can obtain the bracket \eqref{liebra} via the identification
$P_a=i\sqrt{|\Lambda|}J_a$, which yields the Lie algebra
$\mathfrak{sl}(2,\CC)$, realised as the complexification of its
compact real form $\mathfrak{su}(2)$ and its normal real form
$\mathfrak{sl}(2,\RR)$ for Euclidean and Lorentzian signature,
respectively . For $\Lambda>0$, one can introduce an alternative set
of generators
\begin{align}
J_a^\pm=\tfrac{1}{2}(J_a\pm\tfrac{1}{\sqrt{\Lambda}}P_a),
\end{align}
in terms of which the Lie bracket takes the form of a direct sum
\begin{align}
[J_a^\pm,J_b^\pm]=\epsilon_{abc} J^c_\pm,\qquad [J_a^\pm,J_b^\mp]=0.
\end{align}
Hence, depending on the signature and the sign of $\Lambda$, the Lie
algebra $\gothh_\Lambda$ is given by
\begin{align}
\label{liealgs}
\gothh_\Lambda^E=\begin{cases} \mathfrak{iso}(3)
&\Lambda=0\\\mathfrak{su}(2)\oplus \mathfrak{su}(2) &\Lambda>0\\
\mathfrak{sl}(2,\CC)&\Lambda<0,
\end{cases}\qquad\qquad
\gothh_\Lambda^L=\begin{cases} \mathfrak{iso}(2,1)
&\Lambda=0\\\mathfrak{sl}(2,\RR)\oplus \mathfrak{sl}(2,\RR) &\Lambda>0\\
\mathfrak{sl}(2,\CC) &\Lambda<0.
\end{cases}
\end{align}
For all values of $\Lambda$ and both signatures, the space of
$\Ad$-invariant symmetric bilinear forms on $\gothh_\Lambda$ is
two dimensional and a basis is given by the forms
$t,s:\gothh_\Lambda\times\gothh_\Lambda\rightarrow\RR$ defined via
\begin{align}
\label{pair} &t( J_a,J_b)=0, & &t(P_a,P_b)=0, &
&t( J_a,P_b)=\eta_{ab},\\
\label{othform} &s(J_a,J_b)=\eta_{ab}, & &s(J_a,P_b)=0, &
&s(P_a,P_b)=\Lambda\eta_{ab}.
\end{align}

It was shown in \cite{ich3} that the Lie algebras
$\gothh_\Lambda$ can be described in a common framework by
identifying them with the three-dimensional rotation and Lorentz
algebra over a commutative ring defined as follows.

\begin{definition} (Ring $R_\Lambda$ \cite{ich3})
\label{ringdef}

$R_\Lambda=(\RR^2,+,\cdot)$ is the commutative ring obtained from
$\RR^2$ with the usual  addition by defining the
$\Lambda$-dependent multiplication law
\begin{align}
\label{rmult} (a,b)\cdot(c,d)=(ac+\Lambda bd\,,\,
ad+bc)\qquad\forall a,b,c,d\in\RR.
\end{align}
In the following we parametrise elements of $R_\Lambda$ in terms of
a formal parameter $\tl$ as $a+\tl b$, $a,b\in\RR$ and denote by
$\text{Im}_\tl$, $\text{Re}_\tl$ their components
\begin{align}
\label{imredef} \text{Re}_\tl(a+\tl b)=a\qquad \text{Im}_\tl(a+\tl
b)=b\qquad \forall a,b\in\RR.
\end{align}
 We define a
$\RR$-linear involution $\mbox{}^*: \; R_\Lambda\rightarrow
R_\Lambda$, in the following referred to as conjugation, via
\begin{align}
\label{ringconj} (a+\tl b)^*=a-\tl b.
\end{align}
\end{definition}
$R_\Lambda$ is actually more than a ring: it is an algebra over $\RR$
since multiplication by real numbers is also defined. However, since multiplication by real
numbers can  be seen as a special case of multiplication in $R_\Lambda$ we do not emphasise
this aspect, and continue to refer to $R_\Lambda $ as a ring.
Note that the multiplication law \eqref{rmult} follows from  the formal
relation  $\tl^2=\Lambda$. The ring $R_\Lambda$ can therefore  be
viewed as a generalisation of the complex numbers.
For $\Lambda<0$,
the relation $\tl=i\sqrt{|\Lambda|}$ identifies the ring $R_\Lambda$
with the field $\CC$. For $\Lambda=0$ and $\Lambda>0$  the ring
$R_\Lambda$  has zero divisors and can be identified with the dual numbers \cite{dual} and split complex or hyperbolic
numbers \cite{Cockle2},
respectively.
In the case of $\Lambda=0$ the zero
divisors are the elements of the form $\tl a$, $a\in\RR$ satisfying
\begin{align}
\tl a\cdot \tl b=0\qquad\forall a,b\in\RR,
\end{align}
for $\Lambda>0$, the zero divisors are the elements of the form
$\frac{a}{2}(1\pm \tfrac{\tl}{\sqrt{\Lambda}})$, $a\in\RR$, which
satisfy
\begin{align}
\label{posmult} \tfrac{a}{2}(1\pm
\tfrac{\tl}{\sqrt{\Lambda}})\cdot\tfrac{b}{2}(1\pm
\tfrac{\tl}{\sqrt{\Lambda}})=\tfrac{ab}{2}(1\pm
\tfrac{\tl}{\sqrt{\Lambda}})\qquad\tfrac{a}{2}(1+
\tfrac{\tl}{\sqrt{\Lambda}})\cdot \tfrac{b}{2}(1-
\tfrac{\tl}{\sqrt{\Lambda}})=0\qquad\forall a,b\in\RR.
\end{align}

\begin{lemma} \cite{ich3}
\label{lielem}
Consider the three-dimensional rotation and Lorentz algebra with
generators $J_a$, $a=0,1,2$, and with Lie bracket and Killing form
given by \eqref{lorbr}. Extend Lie bracket and Killing form
bilinearly to $R_\Lambda$. With the identification
\begin{align}
\label{ident} P_a=\tl J_a
\end{align}
one recovers the Lie bracket \eqref{liebra} and the $\Ad$-invariant
symmetric bilinear forms \eqref{pair},\eqref{othform} as the real
and $\tl$ component of $\kappa$.
\end{lemma}

\section{The Lie groups of 3d gravity}
\label{groups}
\subsection{Quaternionic structure}

The local isometry groups arising in 3d gravity
are  obtained by exponentiating the  Lie algebras
$\gothh_\Lambda^{E,L}$ \eqref{liealgs}. The fact that these  can be viewed as Lie algebras over
the ring $R_\Lambda$ does not, by itself,  guarantee that the
corresponding Lie groups inherit some kind of algebraic
structure over the ring $R_\Lambda$. In this section we shall explain
that this does, however, happen for the isometry groups of 3d
gravity.
The basic reason for this is
best explained in the context of the Clifford algebras. Recall \cite{LM}
that the Clifford algebra $C\ell(V,\eta)$ associated to a real, $n$-dimensional
vector space $V$ with inner product $\eta$ of signature $(r,s)$ is
 the  associative
algebra  over $\RR$ generated by the elements  of an orthonormal
basis $\{e_0,\ldots,e_{n-1}\}$  of $V$ subject to the relations \bea
\label{cliff} e_ae_b+e_be_a=-2\eta(e_a,e_b)1. \eea The Clifford
algebra contains, as subsets, the original vector space $V$, the
double cover Spin$(r,s)$ of the identity component of  its isometry
group $SO(r,s)$,  and the Lie algebra $so(r,s)$, the latter being
realised as the span of elements $e_ae_b$, with $a\neq b$. As
explained, for example,  in \cite{WW}, the group Spin$(r,s)$ is
realised as a certain  subset  of elements in $C\ell(V,\eta)$
obeying an algebraic condition. When  tensoring $C\ell(V,\eta)$ with
the ring $R_\Lambda$ one obtains, in particular, the Lie algebra
$so(r,s)\otimes R_\Lambda$ as the $R_\Lambda$-span of the degree two
elements. We shall now show that, at least in three dimensions,
 one also  obtains   
corresponding Lie groups by simply interpreting
the algebraic constraint defining Spin$(r,s)$ as an equation in  $R_\Lambda$.
We have found it convenient to express our argument
in the language of quaternions, which exploits the
identification of the degree one  Clifford elements $e_a$ with the degree two Clifford
elements $\frac 1 2 \epsilon_{abc}e_be_c$ in three dimensions.
While this identification, and hence the quaternionic
language, are only possible in three dimensions, the corresponding construction
in the Clifford algebra seems to be possible in any dimension.

\begin{definition} ((Split) Quaternions)
The set of quaternions $\HH^E$ is the associative algebra over
$\RR$  generated by 
elements $e_a$, $a=0,1,2$,  and the identity element $1$ subject to the
relations
\begin{align} \label{quat} e_ae_b=-\eta^E_{ab}\,1 +\epsilon_{abc}e_c,
\end{align}
where $\eta^E$ denotes the three-dimensional Euclidean metric.

The set of split quaternions $\HH^L$ is the associative algebra
over $\RR$ generated by the  elements
 $e_a$, $a=0,1,2$, and the identity element subject to the
 corresponding relations for the three-dimensional Minkowski metric
 \begin{align} \label{quatt} e_ae_b=-\eta^L_{ab}\,1
+\epsilon_{abc}e^c. \end{align}
\end{definition}

Quaternions are discussed in many textbooks on linear algebra,
for example \cite{KM}. The Lorentzian version, called split (or co- or para-) quaternions
was introduced in \cite{Cockle1}, and is
less commonly discussed, but \cite{Klingenberg} contains a detailed
and elementary treatment.
Elements of $\HH^E$ and $\HH^L$ can be parametrised as
\begin{align}
\label{expande} q=q_3\,1+\bq\cd\be=q_3\,1+q^ae_a,  \qquad q^0,q^1,q^2,
q_3\in\RR,
\end{align}
and multiplication is defined by bi-linear extension of
\eqref{quat} and, respectively, \eqref{quatt}. We will mostly omit the identity quaternion $1$ in
the following and write $q=q_3+\bq\cd\be$.
The algebra of (split)
quaternions is equipped with
 an $\RR$-linear conjugation  defined by \bea \label{conj} \bar
q=q_3-q^a e_a.\eea From the identity
\begin{align}
q\bar q=q_3^2 +\bq^2\quad \text{with}\quad \bq=(q^0,q^1,q^2)
\end{align}
it follows that the quaternions $\HH^E$ form a division algebra,
i.~e.~every quaternion $q\neq0$ has a multiplicative inverse. This does not hold for the split quaternions,
but in both cases the set
of unit (split) quaternions satisfying $q\bar q=1$ forms a group:
\begin{align}
&\HH^E_1\!=\!\{q\in\HH^E\,|\, q\bar q=1\}\!=\!\{ q=q_3+q^ae^E_a\,|\,
q_3^2+ q_0^2+q_1^2+q_2^2=1\}\!\cong\!
Spin(3)\!\cong\! SU(2),\\
&\HH^L_1\!=\!\{q\in\HH^L\,|\, q\bar q=1\}\!=\!\{ q=q_3+q^ae^L_a\,|\,
q_3^2+
q_0^2-q_1^2-q_2^2=1\}\!\cong\!Spin(2,1)\!\cong\!SL(2,\RR)\nonumber.
\end{align}
For the corresponding Lie algebras we have
\begin{align}
&\gothh^E:=\{q=q^ae^E_a\in\HH^E\,|\, q^a\in\RR
\}\cong\mathfrak{so}(3)\cong\mathfrak{su}(2),\\
&\gothh^L:=\{q=q^ae^L_a\in\HH^L\,|\, q^a\in\RR
\}\cong\mathfrak{so}(2,1)\cong\mathfrak{sl}(2,\RR).
\end{align}

These isomorphisms can be made explicit via representations. In the
Euclidean case, a representation of the algebra $\HH^E$ is given by
\begin{align}
 \label{so3rep}
\rho_E(1)=\bpm 1 & 0 \\ 0 &1 \epm, \qquad
\rho_E(e^E_a)= -i\sigma_a, \end{align} where
$\sigma_a$ are the Pauli matrices. For Lorentzian signature, two
 representations are relevant. The first induces
a group isomorphism $\HH^L_1\rightarrow SL(2,\RR)$ and is given by
\begin{align} \label{so21rep}
\rho_L(1)=\bpm 1 & 0 \\ 0 &1 \epm,\quad
\rho_L(e^L_0) =\bpm 0 & 1 \\ -1 & 0 \epm,
\quad \rho_L(e^L_1) =\bpm 1 & 0 \\ 0  & -1 \epm, \quad
\rho_L(e^L_2) =\bpm 0 & 1 \\ 1 & 0 \epm, \end{align} while the
second identifies $\HH^L_1$ with $SU(1,1)$ and takes the form
 \begin{align} \label{su11rep}
\tilde \rho_L(1)=\bpm 1 & 0 \\ 0 &1 \epm,\quad
\tilde\rho_L(e^L_0) =\bpm i & 0 \\ 0 & -i \epm,
\quad \tilde\rho_L(e^L_1) =\bpm 0 & i \\ -i  & 0 \epm, \quad
\tilde\rho_L(e^L_2) =\bpm 0 & 1
\\ 1 & 0 \epm. \end{align}

To show how quaternions  give a unified description of the local isometry groups in 3d
gravity
we need to consider quaternions over the commutative ring
$R_\Lambda$. Formally, this means we consider the tensor product
\bea
\HH^{E,L}(R_\Lambda):=\HH^{E,L}\otimes_\RR R_\Lambda.
\eea
Elements can be parametrised according to
\begin{align}
\label{genquat} g=q_3+\tl k_3+\left(\bq+\tl \bk\right)\cd
\be,\qquad q_3,k_3\in\RR, \bq,\bk\in\RR^3,
\end{align}
and it is easy to check that $\HH^{E,L}(R_\Lambda)$ is an algebra over $\RR$.

Both the Lie groups  and the Lie algebras arising in 3d gravity can
be realised as subsets of $\HH^{E,L}(R_\Lambda)$. In order to state this claim
precisely,
 we introduce the projection operator
\begin{align}
\label{project}
\Pi: \HH^{E,L}(R_\Lambda)\rightarrow \HH^{E,L}, \quad
\Pi: p_3+\bp\cd \be \mapsto p_3.
\end{align}

\begin{theorem}
The local isometry groups  in 3d gravity are isomorphic to the multiplicative group
\bea
\HH_1^{E,L}(R_\Lambda):=\{g\in \HH^{E,L}(R_\Lambda)|g\bar g=1\}
\eea
of
unit (split) quaternions over the commutative ring $R_\Lambda$.
We have the following identifications:
\begin{align}
\label{groupident} &\HH_1^E(R_{\Lambda>0})\cong SU(2)\times SU(2),  &
&\HH_1^L(R_{\Lambda>0})\cong SL(2,\RR)\times SL(2,\RR),\\
&\HH_1^E(R_{\Lambda=0})\cong SU(2)\ltimes\RR^3, &
&\HH_1^L(R_{\Lambda=0})\cong SL(2,\RR)\ltimes \RR^3,\nonumber\\
&\HH_1^E(R_{\Lambda<0})\cong SL(2,\CC), &
&\HH_1^L(R_{\Lambda<0})\cong SL(2,\CC).\nonumber
\end{align}
The Lie algebras listed in \eqref{liealgs}  are realised as the set of (split)
quaternions over $R_\Lambda$ with vanishing unit component. In terms of the
projection \eqref{projdef},
\bea
\gothh^{E,L}_\Lambda= \{g\in \HH^{E,L}(R_\Lambda)|\Pi(g)=0\}.
\eea
\end{theorem}

{\bf Notational convention}: \; In the following we will omit the superscript $E$ or $L$
on $\HH$ if the statement being made is valid for either choice of signature.

{\bf Proof:}\; It is easy to check that $\HH_1(R_\Lambda)$ is a
group under multiplication. For Euclidean and Lorentzian signature
and $\Lambda<0$  the identities \eqref{groupident} can be verified
directly by setting $\tl=i\sqrt{|\Lambda|}$ and extending the
representations \eqref{so3rep}, \eqref{so21rep}, \eqref{su11rep} of
the quaternions linearly to $\CC$, which defines three algebra
isomorphisms $\rho_E,\rho_L,\tilde\rho_L:
\HH(R_{\Lambda<0})\rightarrow M(2,\CC)$. A general quaternion
$g\in\HH(R_{\Lambda})$ parametrised as in \eqref{genquat} is a unit
quaternion over $R_\Lambda$ if and only if the parameters $q_3,k_3,
q^a,k^a$ satisfy the conditions
\begin{align}
\label{uquatcond} &q_3k_3+\bq\bk=0, & &q_3^2+\Lambda
k_3^2+\bq^2+\Lambda\bk^2=1.
\end{align}
Using formulas \eqref{so3rep}, \eqref{so21rep}, \eqref{su11rep} it
can be shown by direct calculation that this is equivalent to the
conditions $\det(\rho_E(g))=1$, $\det(\rho_L(g))=1$, $\det(\tilde
\rho_L(g))=1$. This implies that the algebra isomorphisms
$\rho_E,\rho_L,\tilde\rho_L: \HH(R_{\Lambda<0})\rightarrow M(2,\CC)$
restrict to group isomorphisms from $\HH_1(R_{\Lambda<0})$ to
$SL(2,\CC)$.

To prove the corresponding statements for the case of $\Lambda>0$,
we note that we can express a general element
$g\in\HH(R_{\Lambda>0})$ parametrised as in \eqref{genquat} as
\bea
g&=&\tfrac{1}{2}(1+ \tfrac{\tl}{\sqrt{\Lambda}})u_+(g)+\tfrac{1}{2}(1-
\tfrac{\tl}{\sqrt{\Lambda}})u_-(g), \;\;\mbox{with}\\
u_\pm(g)&=&\left(q_3+q^ae_a\right)\pm\sqrt{\Lambda}\left(k_3+k^ae_a\right). \nonumber
\eea
A direct calculation then shows that $g$ satisfies the condition
\eqref{uquatcond} if and only if both elements $u_+(g),
u_-(g)\in\HH$ are unit quaternions $u_+(g),u_-(g)\in\HH_1$.
Moreover, identity \eqref{posmult} implies
\begin{align}
u_\pm(gh)=u_\pm(g)\cdot u_\pm(h)\qquad\forall
g,h\in\HH_1(R_{\Lambda>0}).
\end{align}
By setting
\begin{align} \Phi(g)=\left(\rho(u_+(g))\,,\,
\rho(u_-(g))\right),
\end{align}
where $\rho$ is one of the representations \eqref{so3rep},
\eqref{so21rep}, \eqref{su11rep}, we then
 obtain group isomorphisms $\Phi_E:
\HH_1^E(R_{\Lambda>0})\rightarrow SU(2)\times SU(2)$, $\Phi_L:
\HH_1^L(R_{\Lambda>0})\rightarrow SL(2,\RR)\times SL(2,\RR)$,
$\tilde\Phi_L: \HH_1^L(R_{\Lambda>0})\rightarrow SU(1,1)\times
SU(1,1)$.

When $\Lambda=0$, the
second  condition in \eqref{uquatcond} reduces to the requirement that
the element $q=q_3+q^ae_a\in\HH$ is a unit quaternion. Moreover, we
note that elements of the form
\begin{align}
\label{r3param}
(1+\tl v^ae_a)\cdot q, \qquad \bv\in\RR^3, q\in\HH_1
\end{align}
are unit quaternions over $R_0$ and that any element
$g\in\HH_1(R_0)$ can be expressed uniquely as
\begin{align}
&g=\left(1+\tl v^a(g)e_a\right)\cdot u(g) & &\text{with} &
&u(g)=q_3+q^ae_a,\\
& & & & &v^a(g)=q_3 k^a-k_3q^a+\epsilon^{abc}q_bk_c\nonumber.
\end{align}
As the multiplication relations \eqref{quat},\eqref{quatt} imply
\begin{align}
u(gh)=u(g)\cdot u(h),\qquad v^a(gh)e_a=v^a(g)e_a+u(g)
v^b(h)e_bu(g)^\inv\qquad\forall g,h\in\HH_1(R_0),
\end{align}
the definition
\begin{align}
\Phi(g)=(\rho(u(g)), \bv(g)),
\end{align}
with $\rho$ given by \eqref{so3rep}, \eqref{so21rep} or
\eqref{su11rep},  gives rise to group isomorphisms $\Phi_E:
\HH_1^E(R_{0})\rightarrow SU(2)\ltimes \RR^3$, $\Phi_L:
\HH_1^L(R_{0})\rightarrow SL(2,\RR)\ltimes \RR^3$, $\tilde\Phi_L:
\HH_1^L(R_{0})\rightarrow SU(1,1)\ltimes \RR^3$.\hfill $\Box$

\subsection{Involutions}
\label{invol}

The  quaternionic conjugation
$\bar{\mbox{}}$ \, defined in \eqref{conj} can be extended $R_\Lambda$-linearly
to an involution
\bea
\bar{\mbox{}}:\HH(R_\Lambda) \rightarrow \HH(R_\Lambda), \quad
(q_3+\theta k_3) +(\bq+\theta\bk)\cd\be \mapsto
(q_3+\theta k_3) -(\bq+\theta\bk)\cd\be.
\eea
Similarly, the conjugation $\mbox{}^*$ defined in \eqref{ringconj}
 can be extended to
an involution \bea {\mbox{}^*}:\HH(R_\Lambda) \rightarrow
\HH(R_\Lambda), \quad (q_3+\theta k_3) +(\bq+\theta\bk)\cd\be
\mapsto (q_3-\theta k_3) +(\bq-\theta\bk)\cd\be. \eea We also often
need to consider the combination of the two involutions $\bar{}$ and
${}^*$, and define \bea \label{circconj} g^\circ = \bar{g}^* \eea
for $g\in \HH(R_\Lambda)$. The involution ${}^\circ$ generalises the
notion of taking the adjoint of a matrix. Setting
$\tl=i\sqrt{|\Lambda|}$ for $\Lambda<0$, using the representation
\eqref{so3rep} and extending it linearly to $\HH(R_\Lambda)$, one
recovers the usual Hermitian conjugation
${}^\dag:\;M(2,\CC)\rightarrow M(2,\CC)$ . Doing the same with the
representations \eqref{so21rep}, \eqref{su11rep} yields
\begin{align}
&\bpm a & b \\ c  & d \epm^\circ = \bpm  \bar d & -\bar b \\ -\bar c
& \bar a \epm\qquad\text{for representation \eqref{so21rep}},\\
& \bpm a & b \\
c & d \epm^\circ = \bpm  \bar a & -\bar c \\ -\bar b & \bar d \epm
\qquad\text{for representation \eqref{su11rep}},
\end{align}
which are the usual anti-algebra automorphisms that characterise
 $SL(2,\RR)$ and $SU(1,1)$, respectively,
via the condition $g^\circ=g^\inv$.

Generally, the subsets of $\HH_1(R_\Lambda)$ defined by the
conditions $g^\circ=g$ and $g^\circ=g^{\inv}$ play an important role
in the following. The set of elements satisfying $g^\circ = g^\inv$
is simply the group of unit quaternions:
\begin{align}
\label{lorchar}
 \HH_1=\{ q\in \HH_1(R_\Lambda)\;|\; q^*=q\}=\{ q\in
\HH_1(R_\Lambda)\;|\; q^\circ=q^\inv\}.
\end{align}
The set
\bea
\label{Wdef}
W=\{w\in \HH_1(R_\Lambda)|w=w^\circ\}
\eea
provides a unified description of the model spacetimes arising in 3d gravity.
To see this, parametrise elements as
\bea
\label{wpara}
w = w_3 +\theta \bw\cd \be,
\eea
for $w_0,w_1,w_2,w_3\in \RR$  satisfying the constraint $w_3^2+ \Lambda\bw^2 =1$;
 in this parametrisation
we therefore necessarily have \bea \Lambda\bw^2\leq 1. \eea The
manifolds
 \bea \label{Hlamdef} W_\Lambda=\{(w_3,\bw)\in
\RR^4|w_3^2+\Lambda \bw^2=1\} \eea
parametrising the elements
\eqref{wpara} are isomorphic to various classical geometries for
different choices of $\Lambda$ and the signature $E$ or $L$.
We again suppress the signature label if we refer to either case, and attach
it if we refer to a specific signature.
 In the
Euclidean case and $\Lambda <0$ we have, with $\Lambda =-1$ for
definiteness, \bea W_{-1}^E=\{(w_3,\bw)\in
\RR^4|w_3^2-w_0^2-w_1^2-w_2^2=1 \}, \eea which is  the two-sheeted
hyperboloid embedded in 3+1-dimensional Minkowski space, i.e.
isomorphic to two copies of 3-dimensional  hyperbolic space. For
Euclidean signature and $\Lambda>0$, we obtain the three-sphere
embedded in four-dimensional Euclidean space \bea
W_{1}^E=\{(w_3,\bw)\in \RR^4|w_3^2+w_0^2+w_1^2+w_2^2=1 \}, \eea and
for Euclidean signature with $\Lambda=0$ two copies of
three-dimensional Euclidean space as hyperplanes in four-dimensional
Euclidean space \bea W_{0}^E=\{(w_3,\bw)\in \RR^4|w_3^2=1 \}. \eea

In the Lorentzian case with $\Lambda <0$  we have \bea
W_{-1}^L=\{(w_3,\bw)\in \RR^4|w_3^2-w_0^2+w_1^2+w_2^2=1 \}, \eea
which is the single-sheeted hyperboloid, again
embedded in (3+1)-dimensional Minkowski space; this space is
isomorphic to the double cover of  (2+1)-dimensional  de Sitter space. In the Lorentzian
case with $\Lambda > 0$  we have \bea W_{1}^L=\{(w_3,\bw)\in
\RR^4|w_3^2+w_0^2-w_1^2-w_2^2=1 \}, \eea which is isomorphic to the
double cover of
(2+1)-dimensional
anti-de Sitter space\footnote{Some authors refer to the spaces
 $W_{-1}^L$  and $W_1^L$ as de Sitter and anti-de Sitter space; we use the conventions of \cite{BB}.}.
Finally, for $\Lambda=0$ and Lorentzian
signature, we obtain \bea W_{0}^L=\{(w_3,\bw)\in \RR^4|w_3^2=1 \},
\eea which is simply two copies of (2+1)-dimensional Minkowski space
realised as hyperplanes inside (3+1)-dimensional Minkowski space.
Thus, for each signature and value of the cosmological constant, we obtain
(in some cases two copies of) the corresponding model space of 3d gravity.

For each signature and value of $\Lambda$, elements $g$ of
the group $\HH_1(R_\Lambda)$ act on the set $W$  via
\bea
\Iso (g) : \,W\rightarrow W, \qquad w \mapsto g w g^\circ.
\eea
Geometrically, this is the natural  action of each of the 
local isometry groups \eqref{groupident} arising in 3d gravity on  the (double covers of) 
the corresponding model spacetimes.
In particular we therefore obtain actions of  the unit (split) quaternions
$\HH_1$ on $W$, which we will need later in this paper. For $v\in\HH_1$,
the action 
\bea
\label{adact}
\Iso(v): W\rightarrow W, \quad w \mapsto v w\bar{v}
\eea
is the natural action of the 
rotation group $SO(3)$ or
the orthochronous Lorentz group $SO^+(2,1)$ on the model spacetimes. 
In the following we shall use the notation $\Iso(v)$ for the
action  of $v\in \HH_1$ on both the set $W\subset \HH(R_\Lambda)$ and
the spaces $W_\Lambda\subset \RR^4$ used to parametrise $W$.

\section{The classical double}

In this section we show how to equip $\HH_1(R_\Lambda)$ with the
Poisson-Lie structure of a classical double, and study its group structure in detail.
 Our construction works
for arbitrary values of $\Lambda$ in the Lorentzian case, but only
for  $\Lambda\leq 0$ in the Euclidean case. We begin by exhibiting
Lie-bialgebra structures associated to $\gothh_\Lambda$.

\subsection{Bialgebra structures and classical $r$-matrices }
Our results  about the bialgebra structures on $\gothh_\Lambda$ follow
from a purely  Lie-algebraic observation:
\begin{lemma}
\label{decomplem}
Let $\bn=(n^0,n^1,n^2)$ be a vector in $ \RR^3$ satisfying
\bea
\label{nconst}
\bn^2=\eta_{ab}n^an^b=-\Lambda.
\eea
In terms of the generators
 \begin{align}
\label{genexp} J_a=\tfrac{1}{2}e_a, \qquad  S_a=\tfrac{\tl}{2}e_a+\tfrac 1 2 \epsilon_{abc}n^b e^c=P_a+\epsilon_{abc} n^b J^c
\end{align}
of $\gothh_\Lambda$, the Lie brackets then take the form
\begin{align}
\label{JSbrackets}
[J_a,J_b]=\epsilon_{abc}J^c,\qquad
[J_a,S_b]=\epsilon_{abc} S^c+ n_b J_a-\eta_{ab} (n^cJ_c),\qquad
[S_a,S_b]= n_aS_b-n_b S_a.
\end{align}
\end{lemma}

{\bf Proof}: \; This is a  direct calculation using the relations
for the quaternions, the condition  $\bn^2=-\Lambda$ and the
following
 identity for the epsilon-tensor
\bea
\epsilon_{abc}\epsilon^{cde}=\delta^d_a\delta^e_b-\delta^e_a\delta^d_b.\qquad\qquad\Box
\eea

The lemma shows in particular that both $\gothh$ and the span of
$\{S_0,S_1,S_2\}$ form Lie subalgebras of $\gothh_\Lambda$. The
latter depends on the choice of the vector $\bn$. For $\bn=0$, which
requires $\Lambda=0$, it is simply $\RR^3$ with the trivial Lie
bracket. For $\bn\neq 0$, we will denote this Lie algebra by
$\mathfrak{an}(2)_\bn$ or, when the dependence of $\bn$ need not be
emphasised, simply by $\mathfrak{an}(2)$ in the following. The
reason for the notation $\mathfrak{an}(2)$ is that the decomposition
\bea \gothh_\Lambda = \gothh \oplus \mathfrak{an}(2)_\bn \eea
implied by Lemma~\ref{decomplem} generalises the Iwasawa
decomposition of $\mathfrak{sl}(2,\CC)$ into a compact part
$\mathfrak{su}(2)$ and a ``real abelian + nilpotent part''
$\mathfrak{an}(2)$.

To exhibit the structure of $\an(2)_\bn$ more clearly
and to prepare for calculations later in this paper we introduce
new generators. Consider first the case $\Lambda \neq 0$, and pick a vector $\bm$ orthogonal to $\bn$ but otherwise arbitrary.
Let
\bea
\label{nqformula}
N=-\frac {2}{\Lambda } \bn\cd \bS= -\frac 1 \tl \bn\cd \be \qquad \mbox{and} \qquad Q=\bm\cd\bS =
\frac{\tl }{2} \bm\cd\be + \frac 1 2 \bm\wedge\bn\cd\be.
\eea
Using $n_a n^a=-\Lambda$ one checks  furthermore that
\bea
\label{tqformula}
\tl Q = \bm\wedge\bn\cd \bS,
\eea
so that  $\{N,Q,\tl Q\}$ is an alternative basis of the $\an(2)$ Lie subalgebra.
One finds
\bea
\label{nqalg}
N^2=1,\qquad  Q^2=(\tl Q)^2=0, \qquad NQ=-QN=Q.
\eea
so that, in particular, the commutators take the
form
\bea
\label{nqbrack}
[N,Q]=2Q, \quad [N,\tl Q]=2 \tl Q, \quad [Q,\tl Q]=0,
\eea
showing that $\an(2)$ is isomorphic to the Lie algebra of the semi-direct product $\RR\ltimes \RR^2$.

When $\Lambda =0$ and $\bn\cd\bn=0$ with $\bn\neq 0$ (i.e. in the Lorentzian case),
division by $\tl$ is ill-defined, and we need to modify the definition \eqref{nqformula}.
 We introduce a second light-like vector $\tbn $ which satisfies
\bea
\label{tndef}
\tbn\cd\tbn=0, \qquad \bn\cd\tbn =1.
 \eea
Then the vector
\bea
\label{tmdef}
\tilde \bm =\tbn\wedge \bn
\eea
is space-like, with
 $\tilde \bm\cd \tilde \bm=-1$.
Now define the following generators of the $\an(2)$ Lie subalgebra
\bea
\label{nqgens}
N&=&2\tbn\cd\bS = \tl \tbn\cd \be + \tilde\bm\cd \be, \nonumber \\
Q&=&\tilde \bm\cd \bS = \tfrac{\tl}{2}\tilde \bm\cd \be +\tfrac 1 2 \bn \cd \be, \nonumber \\
\tl Q&=&\bn\cd \bS = \tfrac{\tl}{2}\bn \cd \be.
\eea
Like their counterparts in the $\Lambda \neq 0$ case, they satisfy \eqref{nqalg}, and therefore in particular
the commutation relations \eqref{nqbrack}.

The Lie subalgebras $\gothh$ and $ \mathfrak{an}(2)_\bn $  of $ \gothh_\Lambda$ have the additional feature
that they are both isotropic for the non-degenerate, invariant bilinear inner form $t(\cdot,\cdot)$ i.e. $t(X,Y)=0$
if $X,Y\in \gothh$  or $X,Y \in \mathfrak{an}(2)_\bn$.  This means, by definition \cite{CP},  that
 $(\gothh_\Lambda, \gothh, \mathfrak{an}(2)_\bn)$ together with the invariant bilinear form
$t(\cdot,\cdot)$ is  a Manin triple. More generally  we have

\begin{corollary}
\label{mtriple}
 For every vector  $\bn\neq 0$ satisfying \eqref{nconst},
the  triple $(\gothh_\Lambda, \gothh, \mathfrak{an}(2)_\bn)$ together with the invariant bilinear form
$t(\cdot,\cdot)$ defined in \eqref{pair} is a Manin triple.
When $\bn=0$    $(\gothh_0, \gothh, \RR^3)$ is a Manin triple, with the same  invariant bilinear form
$t(\cdot,\cdot)$.
 \end{corollary}

{\bf Proof}: \; The proof for $\bn\neq 0$ follows from the remarks made before the Corollary.
The proof for $\bn=0$ is analogous.
\hfill $\Box$.

As explained in \cite{CP}, this corollary is equivalent, via standard arguments, to the statement  that both $\gothh$ and $\mathfrak{an}(2)_\bn$
have the structure of a Lie-bialgebra, and that they are dual as Lie-bialgebras. More generally
\bea
\label{liedu}
\gothh^*=\left\{ \begin{array}{l l}
 \mathfrak{an}(2)_\bn  & \mbox{for}\quad  \bn\neq 0\\
   \RR^3&\mbox{for} \quad \bn=0
\end{array}
\right.\qquad
\eea
(where ${}^*$ should not be confused with the conjugation in $R_\Lambda$).

Furthermore, the Corollary~\ref{mtriple}  is equivalent to the existence of
a special bi-algebra structure on the Lie algebra $\gothh_\Lambda$:

\begin{corollary}
The Lie algebra $\gothh_\Lambda$ has a canonical Lie-bialgebra structure, called the classical double,  with classical $r$-matrix
\begin{align}
\label{poincr} r= S_a\otimes J_a= P_a\otimes J^a+n^a\epsilon_{abc} J^b\otimes
J^c\;\in \gothh_\Lambda\otimes \gothh_\Lambda,
\end{align}
where $\bn$ is again assumed to satisfy \eqref{nconst}.
In particular $r$ satisfies the classical Yang-Baxter equation.
\end{corollary}

{\bf Proof}: \; This is a standard construction in  the theory of Lie-bialgebras, see e.g. \cite{CP}, Sect. 1.4.
\hfill $\Box$

The $r$-matrix \eqref{poincr} defines the co-commutator of the
Lie-bialgebra $\gothh_\Lambda$. Equivalently, if defines a
commutator on the dual Lie bi-algebra $\gothh_\Lambda^*$. Using the
pairing $t(\cdot,\cdot)$ \eqref{pair} to identify $\gothh_\Lambda^*$
with $\gothh_\Lambda$ as  vector spaces, we thus obtain a second Lie
bracket on $\gothh_\Lambda$, called the dual Lie  bracket in the
following: \bea \label{dualbrackets}
[J_a,J_b]^*=\epsilon_{abc}J^c,\qquad [J_a,S_b]^*=0,\qquad [S_a,S_b]^*=
n_bS_a-n_a S_b. \eea Note that this Lie algebra is simply the direct
sum of the Lie algebras $\mathfrak{su}(2)$ or $\mathfrak{sl}(2,\RR)$
with the Lie algebra $\mathfrak{an}(2)_{\bn}$, $\bn\neq 0$, or
$\RR^3$ for $\bn=0$.

\subsection{Group structure of the classical double and factorisation}
\label{groupfac} Most of the results about Lie-bialgebras have
analogues at the group level, which  we briefly review. A Manin
triple or, equivalently, a classical double with its canonical
$r$-matrix, exponentiates to a Poisson-Lie group which  is (locally)
factorisable \cite{CP}. Specifically, for the Manin triples
$(\gothh_\Lambda, \gothh, \mathfrak{an}(2)_\bn)$ arising in 3d
gravity, we obtain a Poison-Lie structure on each of the  groups
$\HH_1(R_\Lambda)$ and the factorisation of elements in a
neighbourhood of the identity into elements belonging to the
subgroup $\HH_1$  obtained by exponentiating $\gothh$ and the
subgroup $AN(2)_\bn$ obtained by exponentiating
$\mathfrak{an}(2)_\bn$.  As for the Lie algebra we will omit the
label $\bn$ on the group when the dependence need not be stressed.
Since $\gothh$ and $\mathfrak{an}(2)_\bn$ are Lie-bialgebras in
duality \eqref{liedu}, the corresponding Lie groups are Poisson-Lie
groups in duality
\bea
\label{groupduality}
\HH_1^*= \begin{cases}AN(2)_\bn &\mbox{for} \;\;\bn\neq
0\\
\RR^3 & \mbox{for}\;\; \bn=0,\end{cases} \eea where we again stress that $^*$ is
not the conjugation in $R_\Lambda$. Finally, the dual Lie bi-algebra
$\gothh_\Lambda$ with Lie brackets \eqref{dualbrackets}
exponentiates to a Poisson-Lie group which is dual to
$\HH_1(R_\Lambda)$. From the brackets \eqref{dualbrackets} it is
obvious that, as a Lie group, the dual group is a  direct product:
\bea \HH_1(R_\Lambda)^*=\begin{cases} \HH_1\times AN(2)_\bn &
\mbox{for} \;\;\bn\neq 0\\ \HH_1\times\RR^3 &\mbox{for} \;\; \bn=0.\end{cases} \eea

 In this subsection, we are mainly concerned with  group structure of
$\HH_1(R_\Lambda)$, leaving the discussion of its Poisson structure for Sect.~6.
 We will derive explicit expressions for the
factorisation of elements in $\HH_1(R_\Lambda)$,  and show that it is possible to treat all signs of
$\Lambda$ and signatures in a common framework.
The case $\bn=0$ can be seen as a degenerate limit of our construction.
It is less interesting than the generic situation $\bn$, but we will
on occasion  highlight some of its features. We will assume $\bn \neq 0$
unless stated otherwise.

Elements  of $AN(2)$ can be
parametrised in a number of ways, two of which are important for us
in the following. The first is a parametrisation of elements $t\in
AN(2)$ in terms of an unconstrained vector $\bq \in \RR^3$: \bea
\label{anparam} t=\sqrt{1+(\bq\bn)^2/4} +\bq \cd \bS =
\sqrt{1+(\bq\bn)^2/4} +\frac{\tl}{2} \bq\cd\be +\frac 1 2
\bq\wedge\bn\cd \be. \eea
In the limit  $\bn \rightarrow 0$
this expression reduces  to $1+\tfrac{\tl}{2}\bq\cd\be$, which agrees (apart from a factor $\tfrac 1 2  $)
 with the  earlier parametrisation \eqref{r3param} of elements in $\RR^3$.

 The second parametrisation for elements in $AN(2)$ makes use of
the generators \eqref{nqformula} and \eqref{tqformula} for $\Lambda
\neq 0$ and \eqref{nqgens} for $\Lambda=0$. It takes
the form \bea \label{rpar} r(\alpha,z)=(1+zQ)e^{\alpha N}, \eea
where $z \in R_\Lambda$ and $\alpha \in \RR$. To see that this is a
valid parametrisation, we relate it to the earlier parametrisation
\eqref{anparam}, focussing on the case $\Lambda\neq 0$ (the
calculation for $\Lambda =0$ is similar). Using the relation \bea
\label{handy} e^{\alpha N} =\cosh \alpha + \sinh \alpha\,\, N \eea
we deduce \bea r(\alpha,z)=\cosh\alpha + \sinh\alpha \,N
+e^{-\alpha}z \, Q. \eea In particular, using \eqref{nqformula} and
\eqref{tqformula}, and expanding $z=\xi+\tl\eta$  we can therefore
write \begin{align} &r(\alpha,z)= \cosh\alpha
-\frac{2}{\Lambda}\sinh\alpha\; \bn\cd\bS + e^{-\alpha}\xi\;
\bm\cd\bS + e^{-\alpha}\eta\;(\bm\wedge\bn)\cd
\bS\quad\text{for}\quad \Lambda\neq 0.
\end{align} Comparing
with \eqref{anparam} we find the following relation between the
parameters $\bq$ and $(\alpha, z)$: \bea \bq =
-\frac{2}{\Lambda}\sinh\alpha \;\bn+ e^{-\alpha}\xi\;\bm +
e^{-\alpha}\eta\;(\bm\wedge\bn). \eea The parametrisation
\eqref{rpar} is useful for a number of purposes. It makes manifest
the semidirect product structure of $AN(2)$.  The relation \bea
r(\alpha_1,z_1)r(\alpha_2,z_2)=r(\alpha_1+\alpha_2, z_1 +
e^{2\alpha_1} z_2). \eea follows directly from $N Q= - QN=Q$, and
shows that $AN(2)\simeq \RR\ltimes \RR^2$. The main use of the
parametrisation \eqref{rpar} for us is the dressing action of
$AN(2)$ on $\HH_1$, to be discussed in Sect.~\ref{dresssect}

The subgroup $ AN(2)$  is intimately related to
  the subset  $W$ of $\HH(R_\Lambda)$  introduced and discussed in Sect.~\ref{invol},  whose
elements satisfy $w^\circ =w$ and parametrise the associated model
spacetimes. It is clear that for any $g\in \HH_1(R_\Lambda)$,
$g^\circ g\in W$ and $gg^\circ \in W$. This holds in particular for
elements $t\in AN(2)$. We would like to know if, and under which
conditions, the maps \begin{align} \label{Sqd} &\Sq:AN(2)\rightarrow
W, \quad t\mapsto t^\circ t,\\ &\tilde \Sq: AN(2)\rightarrow W, \quad
t\mapsto  t t^\circ \end{align} can be inverted. In discussing these
maps we shall often not distinguish between the subsets $AN(2)$ and
$W$ of $\HH_1(R_\Lambda)$ and the sets $\RR^3$ and $W_\Lambda$
\eqref{Hlamdef} used to parametrise them. Thus, using the
parametrisation \eqref{anparam} for $t\in AN(2)$ in terms of $\bq
\in \RR^3$, the map $\Sq$ can be written explicitly as \bea
\label{Sdef} \Sq:\RR^3\rightarrow W_\Lambda , \quad \bq \mapsto
(1-\frac{\Lambda}{2}\bq^2, \sqrt{1+(\bq\bn)^2/4}\;\bq +\frac 1 2
\bq\wedge (\bq\wedge \bn)). \eea Similarly, the map  $\tilde \Sq$
 takes the form \bea \label{Stdef} \tilde
\Sq:\RR^3\rightarrow W_\Lambda , \quad \bq \mapsto
(1-\frac{\Lambda}{2}\bq^2, \sqrt{1+(\bq\bn)^2/4}\;\bq -\frac 1 2
\bq\wedge (\bq\wedge \bn)). \eea

\begin{lemma}
\label{inverse} The map $\Sq$ of \eqref{Sdef} is injective but not,
 in general,  surjective. Its image  is
\bea W_{\Lambda}^+=\{(w_3,\bw)\in W_\Lambda|w_3+\bw\bn >0\}. \eea
Restricted to this set, the inverse exists and is  given by \bea
\label{Sinv} S^{-1}: W_{\Lambda}^+\rightarrow \RR^3, \quad (w_3,\bw)
\mapsto \left\{
\begin{array}{l l}
 \frac{1}{\sqrt{w_3 +\bw\bn}}\left(\bw +\frac{1-w_3}{\Lambda}\bn\right) & \mbox{if} \quad \Lambda \neq 0 \\
  \frac{1}{\sqrt{1 +\bw\bn}}\left(\bw +\frac{\bw^2}{2}\bn\right) &\mbox{if}\quad \Lambda =0.
\end{array}
\right. \eea An analogous statement holds for the map $\tilde \Sq$,
with $\bn$ replaced by $-\bn$.
\end{lemma}

{\bf Proof}: \; Injectivity of $\Sq$  and the formula \eqref{Sinv}
for the inverse of $\Sq$ can be shown by evaluating $\Sq^\inv \circ
\Sq (\bq)$. With $(w_3,\bw)=\Sq(\bq)$ one checks that \bea
\label{normform} w_3+\bw\bn=(\sqrt{1+(\bq\bn)^2/4}+\frac 1 2
\bq\bn)^2, \eea and using this formula it is straightforward to
confirm
 that $\Sq^\inv \circ \Sq (\bq)=\bq$ as required.  It is clear from the expression for $\Sq^\inv$
that it is only defined on $(w_3,\bw)$ if $w_3+\bw\bn>0$. The proof
for $\tilde \Sq$ follows by replacing $\bn\mapsto -\bn$. \hfill
$\Box$

We make two comments on  the formula \eqref{Sinv}. Firstly, note
that the  expression  for $\Lambda =0$ in \eqref{Sinv} also follows
from the formula for $\Lambda \neq 0$ by Taylor expanding
$w_3=\sqrt{1-\Lambda\bw^2}$ in powers of $\Lambda$ and taking the
limit $\Lambda \rightarrow 0$. Secondly, the condition
$w_3+\bw\bn>0$ can be interpreted geometrically by saying that it
removes a part of the space $W_\Lambda$, i.~e.~the model spacetime,
but the details depend on the value of $\Lambda$, the signature and
also on the choice of $\bn$. In the Euclidean case, with $\Lambda
<0$, we find, using the Cauchy-Schwarz inequality and
$|\bn|=\sqrt{-\Lambda}$, that \bea |\bw\bn|^2\leq -\Lambda\bw^2<
w_3^2. \eea Hence, both the conditions $w_3\pm \bw\bn>0$ are
automatically fulfilled for all $w_3>0$ (upper sheet of the
two-sheeted hyperboloid) but never for $w_3 <0$. Thus the map $S$
establishes a bijection between $\RR^3$ and the upper sheet of the
two-sheeted hyperboloid in the Euclidean case. In the Lorentzian
cases, the condition for invertibility is harder to interpret
geometrically.

We are now ready to state and prove one of the main results of this paper:

\begin{theorem}
\label{keyth} A given  element $g$ in the gravity Lie groups
$\HH_1(R_\Lambda)$ can be factorised into
\begin{align}
\label{lrfactor1}
 g=u\cdot s, \qquad s\in AN(2), \,
u\in \HH_1,
\end{align}
provided $\Pi\left(g^\circ g(1-\tfrac{\bn\cd\be}{\tl})\right)>0$,
where $\Pi$ is defined as in \eqref{project}. The element $g$ can be
factorised into
\begin{align}
\label{lrfactor2} g =r\cdot v,\qquad r \in AN(2),\, v\in \HH_1,
 \end{align}
provided $\Pi\left(gg^\circ (1+\tfrac{\bn\cd\be}{\tl})\right)>0$.
When they exist, the factors are unique and  given by
 \begin{align}
\label{factform1} &s(g)=\frac{1}{2\norm} \left((1+ g^\circ
g)+\frac{\bn\cd\be}{\tl} (1- g^\circ
g)-\tilde\delta_\Lambda\tfrac{\tl}{2}\, \bn\cd\be \,
\left( \frac{1-g^\circ g}{\tl} \right)^2 \right),\\
\label{factform2}& r(g)=\frac{1}{2\norp} \left( (1+ gg^\circ
)-(1-gg^\circ) \frac{\bn\cd\be}{\tl}
  +\tilde\delta_\Lambda \tfrac{\tl}{2}\, \bn\cd\be\,
 \left( \frac{1-g g^\circ}{\tl} \right)^2 \right), \\
\label{factform3}&u(g)= \frac{1}{2\norm}\left(
(g+g^*)-(g-g^*)\frac{\bn\cd\be}{\tl}
 \right),\\
\label{factform4}&
v(g)=\frac{1}{2\norp}\left((g+g^*)+\frac{\bn\cd\be}{\tl}(g-g^*)
\right),
\end{align}
with  $g$-dependent normalisation factors
\begin{align}
\label{normfacs} &\norp=
\sqrt{\Pi\left(gg^\circ(1+\tfrac{\bn\cd\be}{\tl})\right)}, & &\norm=
\sqrt{\Pi\left(g^\circ g(1-\tfrac{\bn\cd\be}{\tl})\right)},
\end{align}
and $\tilde\delta_\Lambda=1$ for $\Lambda=0$ and
$\tilde\delta_\Lambda=0$ for $\Lambda\neq 0$.
\end{theorem}

The factor $\tfrac{1}{\tl}$ in \eqref{factform1} to \eqref{normfacs}
is defined via $\tfrac{1}{\tl}=\tfrac{\tl}{\Lambda}$ for
$\Lambda\neq0$. For $\Lambda=0$, $\tl$ is a zero divisor of the ring
$R_\Lambda$ and division by $\tl$ is ill-defined. Expressions
\eqref{factform1} to \eqref{normfacs} are nevertheless well-defined
for $\Lambda=0$, since all factors multiplied by $\tfrac{1}{\tl}$
are of the form $\tl t$, where $t\in\HH$. Hence, in a slight abuse
of notation we set for $\Lambda=0$ and $w=\tl t$, $t\in\HH$
\begin{align}
\tfrac{1}{\tl}w=\tfrac{1}{\tl}\left( \tl
t\right)=\text{Im}_\tl(w)=t.
\end{align}
All terms involving division by $\tl$ for $\Lambda=0$ are to be
interpreted in this sense in the following.

{\bf Proof}: \;Consider a general element $g\in\HH_1(R_\Lambda)$
factorised as in \eqref{lrfactor1} and \eqref{lrfactor2}. The fact
that the conjugation operation $\circ$ is an anti-group automorphism
which maps elements of $\HH_1$ to their inverses implies
\begin{align}
\label{factact} g^\circ g=s^\circ s=v^\inv r^\circ r v,\qquad
gg^\circ=rr^\circ=uss^\circ u^\inv.
\end{align}
Since both $g^\circ g$ and $gg^\circ$ are  elements of the space $W$
\eqref{Wdef} we can apply Lemma~\ref{inverse} to find \bea
\label{tisit} s=\Sq^\inv(g^\circ g),\qquad r=\tilde\Sq^\inv(g
g^\circ) \eea provided the conditions for the existence of the
inverse are satisfied. The formulas \eqref{factform1} and
\eqref{factform2} are simply a re-writing of \eqref{tisit}, using
the definition \eqref{Sinv}. To see this, focus on \eqref{factform1}
and parametrise \bea \label{interp} g^\circ g = w_3 +\theta
\bw\cd\be. \eea Then \bea \Pi\left(g^\circ
g(1-\tfrac{\bn\cd\be}{\tl})\right) = w_3+\bw\cd\bn, \eea showing
that the condition for existence of the inverse are precisely those
of Lemma~\ref{inverse}. By straightforward calculation one finds
\bea
\frac {1} {2N_-}  \left( (1+ g^\circ g)+\frac{\bn\cd\be}{\tl}
(1- g^\circ g)-\tilde\delta_\Lambda\tfrac{\tl}{2}\, \bn\cd\be \,
\left( \frac{1-g^\circ g}{\tl} \right)^2\right)\hspace{5cm}\\
=\left\{
\begin{array}{l l}
\frac {1}{ 2\sqrt{w_3+\bw\bn} }\left((1+w_3+\bw\bn)+ \tl( \bw +\frac{1-w_3}{\Lambda}\bn)\cd\be +\bw\wedge \bn\cd \be\right) & \mbox{if} \quad \Lambda \neq 0 \\
\frac {1} {2\sqrt{2+\bw\bn} }\left( (2+\bw\bn) +\tl(\bw
+\frac{\bw^2}{2}\bn)\cd \be +\bw\wedge\bn\cd \be\right)
&\mbox{if}\quad \Lambda =0. \nonumber
\end{array}
\right. \eea Using again the relation \eqref{normform} one now
checks that the right-hand side is \bea \label{sparam} s=
\sqrt{1+(\bq\bn)^2/4} +\frac{\tl}{2} \bq\cd\be +\frac 1 2
\bq\wedge\bn\cd \be, \eea with $\bq=\Sq^\inv( (w_3,\bw))$, thus
showing that \eqref{factform1} is equivalent to \eqref{tisit}. The
calculation showing \eqref{factform2} is analogous.

To show the  formula \eqref{factform3} we note that, with $g=us$,
and $u^*=u$ by definition, \bea \label{int} \frac 1 2
(g+g^*)-(g-g^*)\frac{\bn\cd\be}{2\tl} = u \left(\frac 1 2
(s+s^*)-(s-s^*)\frac{\bn\cd\be}{2\tl}\right). \eea Now parametrise
$s$ as in \eqref{sparam}, and compute \bea \label{int1} \frac 1 2
(s+s^*)-(s-s^*)\frac{\bn\cd\be}{2\tl}=\sqrt{1+(\bq\cd\bn)^2/4}+\tfrac
1 2 \bq\bn \eea as well as \bea \label{int2} \Pi(g^\circ
g(1-\tfrac{\bn\cd\be}{\tl}))=\Pi(s^\circ
s(1-\tfrac{\bn\cd\be}{\tl})) =(\sqrt{1+(\bq\cd\bn)^2/4}+\tfrac 1
2\bq\bn )^2. \eea Now \eqref{factform3} follows from \eqref{int}
with the substitutions \eqref{int1} and \eqref{int2}. The proof of
\eqref{factform4} is again entirely analogous. \hfill $\Box$

As a consequence of the geometrical considerations after the proof
of Theorem~\ref{keyth} we have the following
\begin{corollary}
The factorisation of  Theorem  \ref{keyth} is globally defined in
Euclidean case with $\Lambda <0$.
\end{corollary}

The formulas \eqref{factform1}-\eqref{factform4} also hold for the case $\Lambda=0$ and $\bn=0$.
Then   $N_+(g)=N_-(g)=1$ so that the factorisation is globally defined.  The factorisation is   now simply
\begin{align}
u(g)=v(g)=\tfrac{1}{2}(g+g^*),\qquad s(g)=\tfrac{1}{2}(1+g^\circ
g),\qquad r(g)=\tfrac{1}{2}(1+gg^\circ),
\end{align}
showing how to factor an element in the semi-direct product
$\HH_1(R_0) \cong \HH_1 \ltimes \RR^3$
into an
$\HH_1$ and a $\RR^3$ component.

\section{Dressing transformations}
\label{dresssect}

One important use of Theorem \ref{keyth} is  the definition and computation of  dressing transformations.
The interest of dressing transformations  in physics
stems from the fact that their orbits give the symplectic leaves of a Poisson space \cite{setishan,K-S}.
In order to define dressing transformations in the case at hand, note that
the  formulas \eqref{factform1}-\eqref{factform4}  allow one to compute the factors $u \in \HH_1$ and
$s\in AN(2)$, given the factors $v \in \HH_1$ and
$r\in AN(2)$ and conversely. It follows directly from the definition \eqref{lrfactor1} that $u$, when defined,
 is obtained  by a left action
of $r$ on $v$; similarly  the factor $s$, when defined,
is obtained by a right action of $v$ on $r$. Thus we define the  $AN(2)$ left action
\bea
L_r:\HH_1\rightarrow \HH_1,\qquad
v\mapsto L_r(v):=u
\eea
and the $\HH_1$ right action
\bea
R_v:AN(2)\rightarrow AN(2), \qquad
r\mapsto R_v(r):=s.
\eea
In this way we obtain  dressing transformations on the group manifolds $AN(2)$ and $SU(2)$ or
$SL(2,\RR)$. As mentioned at the beginning of Sect.~\ref{groupfac}, the $r$-matrices \eqref{poincr}
give rise to Poisson-Lie structures for both of these groups. We will describe the associated Poisson structures
explicitly in Sect.~6.

In this section we introduce a quaternionic formalism which gives
a simple and unified description of the geometry of dressing transformations
In the case $\Lambda =0$ and  $\bn=0$ the dressing transformations
are very simple: the left action $L_r$ is trivial, and the right action $R_v$ is the adjoint (right) action
of $\HH_1$.  We therefore assume that $\bn\neq 0$ in the following. The geometrical interpretation of the
dressing right action $R_v$ follows directly from Theorem~\ref{keyth}:

\begin{theorem}
In terms of the parametrisation $r=\sqrt{1+(\bq\bn)^2/4} +\bq\cd \bS$, $\bq\in \RR^3$  of $r\in AN(2)$ and the map $\Sq$ defined in
\eqref{Sdef},   the dressing right action of $v\in\HH_1$   is the map
\bea
\label{dressr}
R_v: \bq  \mapsto \Sq^{-1}\left(\Iso(v^{-1})\Sq(\bq)\right),
\eea
where the action $\Iso$ of $v\in \HH_1$ is defined as in \eqref{adact}.
\end{theorem}

{\bf Proof}: The formula \eqref{dressr}  is merely a re-writing of
the first formula  in \eqref{factact}. According to that formula,
one obtains $s^\circ s\in W$  from   $r^\circ r\in W$ by acting with
$\Iso(v^\inv)$. However, in the notation of \eqref{Sqd},
$\Sq(r)=r^\circ r$ and $\Sq^\inv(s^\circ s)=s$. \hfill $\Box$

It is clear from our  discussion about the invertibility of $S$ following the proof of Lemma~\ref{inverse}
that the dressing action $R_v$ is globally defined in the  Euclidean case, but not in the Lorentzian case:
the formula \eqref{dressr} only makes sense
 if $\Iso(v^{-1})(\Sq(q)$ is in the domain of the map $\Sq^{-1}$.  In order to understand the geometry of  dressing
orbits in detail, we use the notation $\Sq(\bq)=(w_3,\bw)$, with the
formulas \eqref{Sdef} for $w_3$ and $\bw$. One then finds that \bea
\label{wq} \bw^2= \bq^2\left(1-\frac{\Lambda}{4}\bq^2\right). \eea
Both $w_3$ and  $\bw^2$ are invariant under the adjoint action
\eqref{adact} of $\HH_1$ on $w$; When $\Lambda\neq 0$  it follows
from $w_3=1-\tfrac{\Lambda}{2}\bq^2 $  that $ \bq^2$ is also
invariant; when $\Lambda =0$ the invariance of $\bq^2$  can be
inferred from \eqref{wq}. It follows that the orbits under the
dressing actions are subsets of the level sets $\bq^2=$const.

In
the Euclidean case it is easy to see that the orbits are actually equal to the level sets. To prove this
we define
\bea
{\cal O}^E_\rho=\{\bq \in \RR^3|\bq^2=\rho\}, \qquad \rho \geq 0,
\eea
and show that the dressing action on ${\cal O}_\rho$ is transitive. Suppose that $\bq, \tilde \bq\in {\cal O}_\rho$.
With $\Sq(\tilde \bq) = (\tilde w_3,\tilde \bw)$ we then have $\bw^2=\tilde \bw^2$ and $w_3=\tilde w_3$. Thus there
exists a  quaternion $v\in \HH_1$ so that $\Iso(v^\inv)w=\tilde w$; this follows from the transitivity of the $SO(3)$
action on the spheres $\bw^2=$const. But then, by definition, $\tilde \bq= R_v(\bq)$, which was to be
shown. Geometrically, the orbits of the dressing action of $\HH_1$ on the vector  $\bq\in \RR^3$ are therefore the
familiar orbits of  $SO(3)$  acting on $\RR^3$: a  point (the origin) or spheres.

The Lorentzian situation is more complicated because the level sets
$\bq^2=$const. are not, in general, connected. We have the trivial orbit consisting
of the origin, the
single-sheeted hyperboloids
 \bea \label{o1}
{\cal O}^L_\rho=\{\bq \in \RR^3|\bq^2=\rho\}, \qquad \rho < 0,
\eea as well as the upper and lower sheet of the two-sheeted
hyperboloid \bea \label{o2} {\cal O}^{L+}_\rho=\{\bq \in
\RR^3|\bq^2=\rho, q_0 >0 \}, \qquad {\cal O}^{L-}_\rho=\{\bq \in
\RR^3|\bq^2=\rho, q_0 <0 \},\qquad \rho >0, \eea and the upper and
lower lightcone \bea \label{o3} {\cal O}^{L+}_0=\{\bq \in
\RR^3|\bq^2=0, q_0 >0 \}, \qquad {\cal O}^{L-}_0=\{\bq \in
\RR^3|\bq^2=0, q_0 <0 \}. \eea Since $SO^+(2,1)$ is a connected
group, each orbit of the dressing action is connected and therefore
must be a subset of one of the components listed above. To check if
the dressing action is transitive on the orbits we again start with
$\bq$ and $\tilde \bq$ in one of the sets \eqref{o1}-\eqref{o3}, and
compute the corresponding images $\Sq(\bq)=(w_3,\bw)$ and $\Sq(\tilde
\bq)=(\tilde w_3,\tilde \bw)$. Again we have $\bw^2=\tilde \bw^2$, but this only
guarantees the existence of a split quaternion $v$ so that
$\Iso(v^\inv)w=\tilde w$ if $\bw^2< 0$ because both lie on the same
single-sheeted hyperboloid in that case.
Note that, because of
\eqref{wq}, this may happen for both $\bq^2<0$ and $\bq^2>0$. Thus
we conclude that the sets \eqref{o1}-\eqref{o3} are indeed dressing
orbits provided the label $\rho$ satisfies $\rho
(1-\tfrac{\Lambda}{4}\rho)<0$. The origin is trivially an orbit; in
all other cases further analysis is need to determine the orbit
geometry.

Next we turn to the dressing left action $L_r$ of $r\in AN(2)$ on $\HH_1$.
Here the cases $\Lambda \neq 0$ and  $\Lambda =0$ with  $\bn \neq 0$ (which is necessarily Lorentzian) require
slightly different conventions and treatments.
We begin with the case $\Lambda \neq 0$ and
recall the definitions \eqref{nqformula} of the quaternion $N$ and $Q$ in terms of the orthogonal  vectors $\bn$
and $\bm$, with $\bn^2=-\Lambda$.
It follows that   $N^2=1$ and that the operators
\bea
\label{projdef}
P=\frac 1 2 (1+N) \qquad \mbox{and} \qquad \bar P  = \frac 1 2 (1-N)
\eea
are projection operators. Furthermore, we define
$ M=\tl \bm\cd \be$ and  $Q=PM$. Then
\bea
Q^\circ = MP= \frac{1}{2}\tl \bm\cd\be - \frac 1 2 \bm\wedge\bn\cd\be,
\eea
 and it is easy to check the following ``projector algebra''
of the elements $P,\bar P, Q, Q^\circ$:
\bea
\label{PQalg}
P^2&=&P,\quad \bar{P}^2=\bar P, \quad P\bar P =\bar P P =0, \nonumber \\
PQ&=&Q, \quad PQ^\circ =0, \quad \bar P Q =0, \quad \bar P Q^\circ  =Q^\circ, \nonumber \\
QP&=&0, \quad Q^\circ  P=Q^\circ ,\quad Q \bar P =Q, \quad Q^\circ \bar{P} =0, \nonumber \\
Q^2&=&0, \quad (Q^\circ)^2=0, \quad Q Q^\circ= -\Lambda m^a m_a P,
\qquad Q^\circ Q =-\Lambda m^a m_a \bar P. \eea In words: $P$ and
$\bar P$ are projection operators when acting from left or right,
and the nilpotent  elements $Q$ and $Q^\circ $ are
eigenstates of the left- and right-projections. The last line in
\eqref{PQalg} suggests the normalisation $
m^am_a=-\frac{1}{\Lambda}$. This can be achieved in the Euclidean
case when $\Lambda < 0$ , and in the Lorentzian case when $\Lambda
>0$; these cases will be referred to collectively as case (I) in the
following. In the Lorentzian case when $\Lambda <0$, however, the
vector $\bn$ is timelike and it is impossible to find a vector
orthogonal $\bn$ which is also timelike. In that case, called (II)
in the following, we choose $m^am_a=\frac{1}{\Lambda}$. To sum up,
we have the relations \bea \quad Q Q^\circ= \pm P,  \qquad Q^\circ Q
=\pm \bar P, \eea where the upper sign refers to (I) and the lower
sign to (II). Finally we note the following properties with respect
to conjugations: \bea P^*=\bar P, \qquad P^\circ = P, \qquad \bar
Q=-Q, \qquad Q^* = -Q^\circ. \eea

An element $v$ of the  subgroup $\HH_1$  of unit (split) quaternions can be parametrised  in terms of the projector algebra elements
$P$ and $Q$ as
\bea
\label{qexpand}
v= x P + y Q +x^* \bar P - y^*Q^\circ,
\eea
with $x, y\in R_\Lambda$ satisfying the constraint
\bea
\label{normagain}
xx^*\pm yy^* = 1,
\eea
where the upper sign refers to case (I), and the lower to case (II).
It is clear that the element $v$ in \eqref{qexpand} satisfies  $v^*=v$,  and easy to check that $v \bar v=1$ is implied by \eqref{qexpand}.
  To see that any element in $\HH_1$ can be written in the form \eqref{qexpand}, write
\bea
x=a+\tl b, \qquad y=c+\tl d
\eea
and find that \eqref{qexpand} is equivalent to
\bea
v= a- b \, \bn\cd \be + c\, (\bm\wedge \bn)\cd \be + \Lambda d \,\bm \cd \be.
\eea
This is  the  expansion of a quaternion in the orthogonal (but not orthonormal)
 basis \linebreak
\hbox{$\{1, \bn\cd \be, (\bm\wedge \bn)\cd \be , \bm \cd \be,\}$; }any element in $\HH_1$ can be written
in this way, provided
$$
a^2-\Lambda b^2\pm (c^2-\Lambda d^2)=1,
$$ which is precisely the condition  \eqref{normagain}.

Putting the formulas of Theorem \ref{keyth} together with  notation of this subsection,
we arrive at the following geometric characterisation of dressing transformations.

\begin{theorem} With the notation of Theorem \ref{keyth}, and assuming that $\Lambda \neq 0$,
the dressing left action of an element $r\in AN(2)$ on an element $v\in \HH_1$ is
\bea
\label{leftact}
L_r(v)= \frac{1}{\norm}\left(rvP +r^*v\bar P\right).
\eea
In terms of the parametrisations \eqref{rpar} of $r$ and \eqref{qexpand} of $v$ the action $L_r$ is the map
\bea
\label{dressv}
(x,y)\mapsto \frac{1}{N_-}
(e^{\alpha}x\mp z e^{-\alpha}y^*,e^{-\alpha}y),
\eea
with
\bea
N_-=\sqrt{ (e^{\alpha}x\mp z e^{-\alpha}y^*)
(e^{\alpha}x^*\mp z^*e^{-\alpha} y )\pm e^{-2\alpha}yy^*}.
\eea
\end{theorem}

Note that the dressing action $L_r$, like the dressing action $R_v$, is globally defined
in the Euclidean case where the factor $N_-$ is always non-zero. In the Lorentzian
case $N_-$ may vanish, so $L_r$  is not globally defined.

{\bf Proof}: \,\, The result \eqref{leftact} is the formula
\eqref{factform3} written in terms of the projection operators $P$
and $\bar P$. Re-writing the parametrisation \eqref{rpar} for $r$ in
the form \bea r(\alpha,z)=e^\alpha P + e^{-\alpha}(\bar P + zQ), \eea
using  \eqref{qexpand} for $v$  and  the relations \eqref{PQalg},
one computes \bea rvP=(e^{\alpha}x\mp z
e^{-\alpha}y^*)P-e^{-\alpha}y^*Q^0. \eea Now observe  that, since
$v^*=v$, the second term in the final expression \eqref{leftact} is
the $*$-conjugate of the first. Thus \bea
L_r(v)=\frac{1}{\norm}\left((e^{\alpha}x\mp e^{-\alpha} z
y^*)P+yQ\;\;+\;\;\mbox{$*$-conjugate}\right). \eea Comparing
with the parametrisation \eqref{qexpand}, and noting that the factor
$1/\norm$ merely ensures that $L_r(v)$ is a unit (split) quaternion,
we conclude that, in terms of the coordinates $x,y\in R_\Lambda$ in
\eqref{qexpand}, $L_r$ is the map \eqref{dressv}. \hfill $\Box$

The formula \eqref{dressv} gives a simple geometrical description of orbits under the dressing transformation $L_r$.
The key observation is that the coordinate $y$ is only multiplied by a real factor. Thus the direction of the
(split) complex number in the (split) complex plane remains unchanged, and the product $yy^*$ is multiplied by
a positive number. The orbit is the set of all $x\in R_\Lambda$ satisfying the constraint
\eqref{normagain} as $yy^*$ is rescaled by an arbitrary positive number. The geometry of this set depends
on the signature and the value of $\Lambda$.

Euclidean signature,\;$\Lambda <0$: This is a much studied case, see e.g. \cite{Majid}.  The constraint $xx^*+yy^*=1$
 defines a three-sphere $S^3_\Lambda$ embedded in $\RR^4$ (and squashed if $\Lambda \neq -1$).  There are two kinds of orbits,
depending on
whether $y=0$ or $y\neq 0$. In the former case the orbits consist of points $(x,0)$ inside $S^3_\Lambda$.
In the second case we automatically have $xx^*<1$, so the orbits consist of discs in the $x$-plane, and are
labelled by the argument of $y$
(which is unchanged by the scaling).

Lorentzian signature,\;$\Lambda > 0$:  The constraint $xx^*+yy^*=1$
 defines the double cover $\widetilde{\text AdS}^3_\Lambda$  of
three-dimensional anti-de Sitter space embedded in $\RR^4$ (and squashed if $\Lambda \neq 1$).
There are three kinds of orbits, depending on
whether $yy^*$ is positive, zero or negative. Since $xx^*=1-yy^*$, the product $xx^*$ is correspondingly less than 1, equal to
1 or bigger than one. Since the equation $xx^*=1$ defines a hyperbola with two branches in the $x$-plane, the
dressing  orbit is the region  between the two branches of the hyperbola when  $yy^*>0$ and the region outside the branches when $yy^* <0$.
When $yy^*=0$ the orbit is the Cartesian  product of the hyperbola (both branches) in the $x$-plane
with the double lightcone in the $y$-plane.

Lorentzian signature \;$\Lambda <0$: The constraint $xx^*-yy^*=1$
again  defines the double cover of
three-dimensional anti-de Sitter space embedded in $\RR^4$ (and squashed if $\Lambda \neq -1$).
This time $yy^*\geq 0$, so there are only two kinds of orbits. If $y=0$ the orbits consist of points $(x,0)$ inside $\widetilde{\text AdS}^3_\Lambda$.
If   $y\neq 0$, we automatically have $xx^*>1$, so in each case the   orbit is the complement of a disc,  and is
 labelled by the argument of $y$.

Turning finally
to the case $\Lambda =0$,  we use the vectors $\tilde\bn$ and $ \tilde\bm$ defined in \eqref{tndef} and \eqref{tmdef}
to parametrise
elements  $v\in \HH_1$ via
\bea
\label{newvpara}
v =a + b \tilde \bm \cd \be + \gamma \bn\cd \be + \tilde \gamma \tbn\cd \be
\eea
in terms of $a,b,\gamma,\tilde \gamma \in \RR$. The condition $v\bar v=1$ is equivalent to
\bea
\label{zeroeq}
a^2-b^2 + 2 \gamma \tilde \gamma =1.
\eea
 Using the  parametrisations
\bea
\label{ragain}
r(\alpha,z)=(1+zQ)e^{\alpha N},
\eea
with $N$ and $Q$ as in \eqref{nqgens},
 we compute $L_r(v)=u$ according to \eqref{factform3}, i.e.
\bea
\label{lrvagain}
L_r(v)= \frac{1}{2\norm}\left((rv +r^*v)- (rv-r^*v)\frac{\bn\cd\be}{\theta}\right).
\eea
With $\Lambda=0$ we have not been able to introduce the analogue of the projector algebra
\eqref{PQalg}, which simplified the calculation in the $\Lambda \neq 0$ case.
However, one can still evaluate \eqref{lrvagain}:

\begin{theorem} In the case $\Lambda=0$, $\bn\neq 0$, the dressing left action $L_r$  of $r\in AN(2)$
 on $v\in \HH_1$ is conveniently expressed in terms of
the parametrisation \eqref{ragain} for $v$
and the $R_0$ coordinates
\bea
\label{xyzero}
x=(a+b)-\tl \gamma, \qquad y= \tilde \gamma +\tl b
\eea
for $v$, written as in \eqref{newvpara}.
In terms of these coordinates, the left action $L_r$
is the map
\bea
\label{dressvzero}
(x,y)\mapsto \frac{1}{N_-}(e^{\alpha}x- z e^{-\alpha}y^*,e^{-\alpha}y),
\eea
where
\bea
N_-=\sqrt{\Rea_\tl((e^{\alpha}x- z e^{-\alpha}y^*)^2)-2\Ima_\tl((e^{\alpha}x- z y^*e^{-\alpha})e^{-\alpha}y)}.
\eea
\end{theorem}

{\bf Proof}:\; This is a direct, somewhat tedious calculation.
The formula for the normalisation factor follows from
$v\bar v= a^2-b^2+2\tilde \gamma \gamma =\Rea_\tl(x^2) -2 \Ima_\tl(xy).$
\hfill $\Box$

The geometry of the dressing orbits is more difficult to understand
in this case. The constraint \eqref{zeroeq} again
defines the double cover of  three dimensional anti-de Sitter space embedded in $\RR^4$, with the coordinates $\gamma$ and $\tilde \gamma$
playing the role of light-cone coordinates. Under the dressing action, the $R_0$ coordinate $y=\tilde \gamma + \theta b$
is rescaled by a positive, real number. In the general case, where $b\neq 0$ and $\tilde \gamma \neq 0$, the equation
\eqref{zeroeq} defines a one-parameter family of parabolas in the $(a,\gamma)$-plane as $y$ is rescaled. In  special
cases, the geometry is simpler. In the trivial case $y=0$ the orbits again consist of a point $(a,\gamma)$. When $\tilde \gamma =0$
and $b\neq 0$ we have $a^2=1+b^2$, so that $a$ ranges over $\RR \setminus [-1,1]$. Since  $\gamma$ is unconstrained the orbit is the
Cartesian
product $(\RR \setminus [-1,1])\times \RR$ in the $(a,\gamma)$-plane. If $b=0$ and $\tilde \gamma \neq 0$ we obtain the family of parabolas
defined by $a^2+2\tilde \gamma\gamma-1=0$ in the $(a,\gamma)$-plane.

\section{Poisson structures associated to the 3d gravity groups}

In this section we consider Poisson structures and Poisson-Lie
structures associated to the Lie groups in 3d gravity, more
precisely the Sklyanin Poisson-Lie structure, the dual Poisson-Lie
structure and the Heisenberg double Poisson structure. We derive
explicit expressions for the Poisson brackets in terms of a set of
natural coordinates derived from their factorisation into subgroups
$\HH_1$ and $AN(2)$ (or $\HH_1$ and $\RR^3$) of and their
identification with the set of unit quaternions $\HH_1(R_\Lambda)$.

A strong motivation for considering these Poisson structures is
their role in the description of the phase space of 3d gravity. It
was shown by Fock and Rosly \cite{FR} that the phase space and
Poisson structure of Chern-Simons theory with gauge group $G$ on
manifolds of topology $\RR\times S_{g,n}$, where $S_{g,n}$ is an
orientable two-surface of genus $g$ with $n$ punctures, can be
described in terms of an auxiliary Poisson structure on the manifold
$G^{n+2g}$. This Poisson structure is defined uniquely in terms of a
classical $r$-matrix for the group $G$. Moreover, it was
demonstrated by Alekseev and Malkin \cite{AMII} that the
contribution of different handles and punctures to this Poisson
structure can be decoupled and related to two well-known Poisson
structures from the theory of the Poisson-Lie groups: each puncture
corresponds to a copy of the dual Poisson-Lie structure on $G$,
while each handle is characterised by a copy of the associated
Heisenberg double Poisson structure. For semidirect product groups
of the form $G\ltimes\gothg^*$ an explicit expression for the
decoupling map and the resulting Poisson structures is given in
\cite{we1,we2}.

In the application to 3d gravity in its formulation as a
Chern-Simons gauge theory with gauge group $\HH_1(R_\Lambda)$, this
implies that Fock and Rosly's auxiliary Poisson structure of the
theory on a general manifold on manifold $\RR\times S_{g,n}$ is
given as the direct product
\begin{align}
\underbrace{\HH_1(R_\Lambda)_D\times\ldots\times
\HH_1(R_\Lambda)_D}_{n\times}\times \underbrace{
{D}_+(\HH_1(R_\Lambda))\times\ldots\times {D}_+(\HH_1(R_\Lambda))
}_{g\times},
\end{align}
where $\HH_1(R_\Lambda)_D$ is the group $\HH_1(R_\Lambda)$ with the
dual Poisson-Lie structure and ${D}_+(\HH_1(R_\Lambda))$ the
manifold $\HH_1(R_\Lambda)\times \HH_1(R_\Lambda)$ with the
Heisenberg double Poisson structure. Hence, determining the dual
Poisson-Lie structure and the associated Heisenberg double Poisson
structure for the Lie groups arising in 3d gravity amounts to
giving a complete description of the phase space and its Poisson
structure for spacetimes of general genus $g$ and with  $n$
punctures representing massive, spinning particles.

In the following we restrict attention to the Lorentzian 3d gravity
with general cosmological constant and the Euclidean case with
$\Lambda\leq 0$, where the Lie groups $\HH_1(R_\Lambda)$ have the
structure of a classical double. The key idea is to combine the factorisation
of the groups derived in Theorem \ref{keyth} and the identification
of the Lie groups in 3d gravity with the set of unit quaternions
$\HH_1(R_\Lambda)$ to obtain a natural set of coordinates, in which
these brackets are of a particularly simple form. More specifically,
for $\bn\neq 0$
we consider the functions $\tilde p^a\in\cif(\HH_1)$, $\tilde
q^a\in\cif(AN(2))$ defined by
\begin{align}
\label{helpcor} &\tilde p^a(u)= -2\Pi\left(u \cdot e^a\right),\qquad
\tilde q^a(s)= -2\text{Im}_\tl \,\Pi\left( s\cdot
e^a\right)\qquad\forall u\in\HH_1,s\in AN(2)
\end{align}
and apply them to the group elements obtained by factorising
$g\in\HH_1(R_\Lambda)$ as $g=u(g)\cdot s(g)$, $u\in\HH_1$, $s\in
AN(2)$ as in \eqref{lrfactor1}. In the case $\bn=0$ we obtain functions
$q^a\in\cif(\RR^3)$ via the same formula as in \eqref{helpcor} but now with
$s\in \RR^3$ (compare our comment after \eqref{anparam}).
 As the factorisation is not global
except for the case $\bn=0$ and the Euclidean case with $\Lambda<0$, the
resulting functions are not defined globally but only on the subset
of factorisable elements of $\HH_1(R_\Lambda)$. We obtain the
following lemma.
\begin{lemma}

Consider the set of factorisable group elements
\begin{align}
\label{factset} F(\bn)=\{ g\in\HH_1(R_\Lambda)\;|\; \Pi\left(g^\circ
g(1-\tfrac{\bn\be}{\tl})\right)>0\}\subset\HH_1(R_\Lambda),
\end{align}
and set, for $g\in F(\bn)$, $a=0,1,2$,
\begin{align}
\label{coordef} p^a(g)=-2\Pi\left(u(g)\cdot e^a\right),\quad
q^a(g)=-2\text{Im}_\tl \,\Pi\left(s(g)\cdot e^a\right),
\end{align}
with $u(g)$, $s(g)$ given by \eqref{factform1}. The coordinate functions $p^a,q^a\in\cif(F(\bn))$, $a=0,1,2$,
 determine group elements
$g\in F(\bn)$ completely up to a minus sign.
\end{lemma}

{\bf Proof:} This follows directly from the uniqueness of the
factorisation and the definition of the coordinate functions
\eqref{helpcor}. Using the parametrisation \eqref{anparam} and the
standard parametrisation of the quaternions, we find that the
coordinate functions $\tilde q^a, \tilde p^a$  are
\begin{align}
\label{tildadef}
\tilde q^a:\;\sqrt{1+(\bq\bn)^2/4}+\bq\cdot\bS\mapsto q^a,\qquad
\tilde p^a:\; p_3+\bp\cdot\bJ\mapsto p^a.
\end{align}
This determines elements of $AN(2)$ (when $\bn \neq 0$) or $\RR^3$ (when $\bn=0$), uniquely and elements of $\HH$
up to a choice of minus sign arising from the relation
$p_3^2=1-\bp^2$. \hfill $\Box$

We will now demonstrate that these coordinates give rise to a
unified description of the Sklyanin and the dual Poisson-Lie
structure on $\HH_1(R_\Lambda)$ in which the Poisson bracket takes a
rather simple form and the structural similarities for the different
signature and signs of $\Lambda$ are readily apparent. In
particular, we have the following theorem.

\begin{theorem} \label{dualskylth}In terms of the coordinate functions
$p^a,q^a\in\cif(F(\bn))$, the Sklyanin bracket $\{\,,\,\}_S$ and the
dual bracket $\{\,,\,\}_D$ on the gravity Lie groups
$\HH_1(R_\Lambda)$ associated to the classical $r$-matrix
\eqref{poincr} are
\begin{align}
\label{dualskyl} &\{p^a,p^b\}_S=p_3( n^a p^b-p^a n^b), &
&\{q^a,q^b\}_S=q_3\epsilon^{abc}q_c, & &\{p^a,q^b\}_S=0,\\
&\{p^a,p^b\}_D=-p_3( n^a p^b-p^a n^b), &
&\{q^a,q^b\}_D=q_3\epsilon^{abc}q_c, &
&\{p^a,q^b\}_D=q_3\epsilon^{abc}p_c+p_3(\bq\bn\,\eta^{ab}-n^aq^b),\nonumber\end{align}
where $q_3,p_3\in\cif(F(\bn))$ are given by
\begin{align}
\label{p3q3def} q_3=\sqrt{1+(\bq\bn)^2/4},\qquad
p_3=\pm\sqrt{1-\bp^2/4}.
\end{align}
The functions
\begin{align}
\label{casfunct} \text{Im}_\tl(\Pi(u\cdot
s))=-\tfrac{1}{4}p_aq^a,\qquad \text{Re}_\tl(\Pi(u\cdot
s))=p_3q_3-\tfrac{1}{4}\epsilon_{abc}n^ap^bq^c
\end{align}
are Casimir functions for the dual bracket $\{\,,\,\}_D$.

\end{theorem}

{\bf Proof:} \;\; The general formula for the Sklyanin-Poisson
structure on a Lie group $G$ with
 associated quasitriangular Lie algebra $\gothg$ can be found for instance in
 \cite{CP}.  The dual  Poisson structure  is introduced and discussed in \cite{setishan}, see also \cite{AMII} and
\cite{AS}.
In terms of a basis
 $X_a$, $a=1,\ldots,\text{dim}(\gothg)$, of $\gothg$ and its classical
 $r$-matrix
 $r=r^{ab}X_a\otimes X_b$, the Sklyanin Poisson-Lie structure and
 its dual on $G$ can be characterised by the Poisson
 bivectors
\begin{align}
&B_S=\tfrac{1}{2}r^{ab}\left(X_a^R\wedge X_b^R- X_a^L\wedge
X_b^L\right),\label{skylb}\\
&B_D=\tfrac{1}{2} r^{ab}\left(X_a^R\wedge X_b^R+ X_a^L\wedge
X_b^L\right)+r^{ab} X_a^R\wedge X_b^L\label{dualb},
\end{align}
where $X_a^L$, $X_a^R$ are the right- and left-invariant vector
fields associated to the generators via
\begin{align}
\label{lrvecfields} X_a^Lf(g)=\frac{d}{dt}|_{t=0}
f(e^{-tX_a}g),\qquad X_a^Rf(g)=\frac{d}{dt}|_{t=0} f(g
e^{tX_a})\qquad\forall g\in G, f\in\cif(G).
\end{align}

To obtain these Poisson structures for the gravity Lie groups
$\HH_1(R_\Lambda)$, we insert the classical $r$-matrix
\eqref{poincr} into \eqref{skylb}, \eqref{dualb}. Since the
coordinate functions $p^a, q^a \in\cif(F(\bn))$ are
invariant, respectively, under right multiplication with elements of
$AN(2)\subset\HH_1(R_\Lambda)$ and left multiplication with elements
of $\HH_1\subset \HH_1(R_\Lambda)$, we find that their Poisson
brackets are
\begin{align}
\label{pbrack1}
&\{p^a,p^b\}_D=-\{p^a,p^b\}_S=\tfrac{1}{2}\left(S_c^Lp^a
J^c_Lp^b-S_c^Lp^b J^c_Lp^a\right),\\
\label{qbrack1}&\{q^a,q^b\}_D=\{q^a,q^b\}_S=\tfrac{1}{2}\left(S_c^Rq^aJ^c_Rq^b-S_c^Rq^bJ^c_Rq^a\right),\\
\label{pqbr1}&\{p^a,q^b\}_D=-\tfrac{1}{2}\left( J_c^Lp^a S^c_Lq^b+
J_c^Rp^a
S^c_Rq^b\right)-J_c^Lp^a S^c_Rq^b,\\
\label{pqsbr1}&\{p^a,q^b\}_S=\tfrac{1}{2}\left( J_c^Lp^a S^c_Lq^b-
J_c^Rp^a S^c_Rq^b\right),
\end{align}
where $J_a^L,S_a^L$ and $J_a^R, S_a^R$ are the right-and
left-invariant vector fields on $\HH_1(R_\Lambda)$ defined as in
\eqref{lrvecfields}.

To evaluate \eqref{pbrack1} to \eqref{pqsbr1}, we need to determine
the action of the vector fields $J_a^L$, $S_a^L$, $J_a^R$, $S_a^R$
on the coordinate functions $p^a,q^a\in\cif(F(\bn))$. The first step
is to determine the action of the left- and right-invariant vector
fields on the groups $\HH_1$ and  $AN(2)$ on the coordinate functions
$\tilde p^a\in\cif(\HH_1)$, $\tilde q^a\in\cif(AN(2))$ and
  $ \tilde q^a\in\cif(\RR^3)$ for $\bn=0$ . Using the
standard parametrisation of the quaternions for $\HH_1$ and the
definition \eqref{helpcor}, we find
\begin{align}
\label{lorvecl} &L_{J_a}\tilde p^b(u)=\frac{d}{dt}|_{t=0}\tilde
p^b(e^{-tJ_a}\cdot
u)=-\eta^{ab}\tilde p_3+\tfrac{1}{2}\epsilon^{abc}\tilde p_c,\\
&R_{J_a}\tilde p^b(u)=\frac{d}{dt}|_{t=0}\tilde p^b(u\cdot
e^{tJ_a})=\eta^{ab} \tilde p_3 +\tfrac{1}{2}\epsilon^{abc}\tilde p_c,
\label{lorvecr}
\end{align}
with
\begin{align}
\tilde p_3(u)=\Pi(u)=\pm\sqrt{1-\tilde \bp^2/4(u)}.
\end{align}
The corresponding calculation for $AN(2)$ and $\RR^3$ using the
 parametrisation \eqref{anparam} and definition \eqref{helpcor} yields
\begin{align} \label{transvecl}
&L_{S_a}\tilde q^b(s)=\frac{d}{dt}|_{t=0}\tilde q^b(e^{-tS_a}\cdot
s)=-
(\tilde q_3-\tfrac{1}{2}\tilde \bq\bn)\,\eta^{ab}-\tfrac{1}{2} n^a \tilde q^b,\\
\label{transvecr}&R_{S_a}\tilde q^b(s)=\frac{d}{dt}|_{t=0}\tilde
q^b(s\cdot e^{tS_a})=(\tilde q_3+\tfrac{1}{2}\tilde
\bq\bn)\,\eta^{ab}-\tfrac{1}{2} n^a \tilde q^b,
\end{align}
with \begin{align} \tilde q_3(s)=\sqrt{1+(\tilde \bq\bn)^2/4(s)}.
\end{align}
This determines the action of the vector fields $J_a^L$, $S_a^R$ on
the coordinate functions $p^a,q^a\in\cif(F(\bn))$:
\begin{align}
\label{lorvecll} &{J_a^L}
p^b=-\eta^{ab}p_3+\tfrac{1}{2}\epsilon^{abc}p_c, & &J_a^L q^b=0,\\
&{S_a^R} q^b=(q_3+\tfrac{1}{2} \bq\bn)\,\eta^{ab}-\tfrac{1}{2} n^a
q^b, & &S_a^Rp^b=0.\label{transvecrr}
\end{align}
To obtain the corresponding expressions for the vector fields
$J_a^R$, $S_a^L$  we combine \eqref{lorvecl}, \eqref{lorvecr} and
\eqref{transvecl}, \eqref{transvecr} with infinitesimal dressing
transformations. Inserting
\begin{align}
v=e^{tJ_a}, \qquad r=q_3+q^a S_a
\end{align}
into \eqref{factform1} and \eqref{factform2}, using the expressions
\eqref{genexp} of the generators $J_a$, $S_a$ in terms of the
quaternions and the multiplication relations \eqref{quat},
\eqref{quatt} and expanding the resulting expressions in $t$ we
obtain
\begin{align}
\label{helppp1} &u=1+t\left(\delta_a^c +\frac{q_a n^c
-(\bq\bn)\delta_a^c}{ q_3+\tfrac{1}{2}\bq\bn}\right)
J_c +O(t^2),\\
&s=\left(1+{t} \frac{\epsilon_{abc}q^bn^d}{
2(q_3+\tfrac{1}{2}\bq\bn)}\right)
r+t\epsilon_a^{\;\;cd}\left(q_d-\frac{\bq^2 n^d}{2(
q_3+\tfrac{1}{2}\bq\bn)}\right)S_c+O(t^2).
\end{align}
Combining \eqref{helppp1} with \eqref{lorvecr} then yields
expressions for the action of the vector field  $J_a^R$ on the
coordinate functions $p^a,q^a$
\begin{align}
&J_a^Rp^b=\frac{q_3-\tfrac{1}{2}\bq\bn}{q_3+\tfrac{1}{2}\bq\bn}\left(
p_3\delta_a^b+\tfrac{1}{2}\epsilon_{a}^{\;\;bc}p_c\right)+\frac{q_a}{q_3+\tfrac{1}{2}\bq\bn}\left(
p_3n^b+\tfrac{1}{2}\epsilon^{bcd}p_cn_d\right),\label{jrp}\\
&J_a^Rq^b=\frac{q_3\epsilon_a^{\;\;bc}q_c-\tfrac{1}{2}q^a\epsilon^{bcd}n_cq_d}{q_3+\tfrac{1}{2}\bq\bn}\label{jrq}.
\end{align}
To determine the action of the vector field $S_a^L$, we insert
\begin{align}
r=e^{-tS_a},\qquad v=p_3+p^aJ_a
\end{align}
into \eqref{factform1}, \eqref{factform3} and expand the resulting
expressions in $t$
\begin{align}
&u=\left(1-t\,\left(p_3(p_an^c-p^cn_a)+\tfrac{1}{2}\bp\bn\,\epsilon_{a}^{\;\;cd}p_d-\tfrac{1}{2}\bp^2\epsilon_a^{\;\;cd}n_d\right)
J_c\right)\cdot v +O(t^2),\label{helppp2}\\
&s=1-t\left(-p_3\epsilon_a^{\;\;cd}p_d+\tfrac{1}{2}p_ap^c+\left(1-\tfrac{1}{2}\bp^2\right)\delta_a^c\right)S_c+O(t^2).
\end{align}
Combining \eqref{helppp2} with \eqref{lorvecl} and applying the
identity
\begin{align}
x^b\epsilon^{acd}y_c x_d-x^a\epsilon^{bcd}y_c
x_d=(\bx\by)\epsilon^{abc}x_c
-(\bx^2)\epsilon^{abc}y_c\qquad\forall\bx,\by\in\RR^3,
\end{align}
we then obtain the action of the vector field $S_a^L$ on the
coordinate functions $p^a$ and $q^a$:
\begin{align}
&S_a^Lp^b=\tfrac{p_3}{2}p^b\epsilon_{acd}p^cn^d+p_3^2
n_ap^b+\tfrac{1}{4}\bp\bn p_ap^b-p_an^b,\label{slp}\\
&S_a^Lq^b=-\left(\left(1-\tfrac{1}{2}\bp^2\right)\eta_{ac}+\tfrac{1}{2}p_ap_c-p_3\epsilon_{acd}p^d\right)\left(\left(q_3-\tfrac{1}{2}\bq\bn\right)\eta^{bc}+\tfrac{1}{2}n^cq^b\right).\label{slq}
\end{align}
Inserting expressions \eqref{lorvecll}, \eqref{transvecrr},
\eqref{jrp}, \eqref{jrq}, \eqref{slp}, \eqref{slq} for the action of
the vector fields $J_a^L$, $J_a^R$, $S_a^L$, $S_a^R$ on the
coordinate functions into formulas \eqref{pbrack1} to \eqref{pqsbr1}
 yields \eqref{dualskyl}. It can then be verified by direct calculation that the functions \eqref{casfunct} Poisson commute with the coordinate functions $p^a$ and  $q^a$ for the dual bracket. \hfill $\Box$

Comparing formulas \eqref{dualskyl} for the Poisson bracket with
formulas \eqref{JSbrackets} and  \eqref{dualbrackets}, we find that, up to
factors $p_3,q_3$ and a minus sign, the dual Poisson bracket
$\{\,,\,\}_D$ and the Sklyanin bracket $\{\;,\;\}_S$
 agree, respectively, with the Lie bracket \eqref{JSbrackets} and dual Lie bracket
\eqref{dualbrackets}. Moreover, the two Casimir functions of the
dual bracket $\{\,,\,\}_D$ are of a particularly simple form and
have a interpretation as the  real and imaginary part of the
group element's unit coefficient in its description as a quaternion
over $R_\Lambda$.

The factorisation of the gravity Lie groups in Theorem \ref{keyth}
does not only give rise to a natural set of coordinates in which the
dual Poisson structure and the Sklyanin bracket on the subset
$F(\bn)\subset \HH_1(R_\Lambda)$ take a particularly simple form,
but also provides useful information about the
Poisson structures  of the subgroups $\HH_1$ and $AN(2)_\bn$ (or
$\RR^3$ when $\bn =0$). As discussed after \eqref{groupduality}
these are mutually dual Poisson-Lie groups. The Sklyanin bracket
$\{\,,\,\}_S$ on the double $\HH_1(R_\Lambda)$ encodes the Poisson
structures of both these subgroups. Furthermore, the dressing 
transformations studied in Sect.~5  determine the symplectic leaves of the
Poisson-Lie groups $\HH_1$ and $AN(2)_\bn$ (respectively $\RR^3$
when $\bn =0$). The results can be summarised as follows.

\begin{theorem}
In terms of the coordinate functions $\tilde p^a$ on $\HH_1$ and $\tilde q^a$ on $(AN(2)_\bn)$
($\RR^3$ when $\bn=0$)
defined in \eqref{tildadef}, the Poisson brackets 
on the Poisson-Lie groups 
$\HH_1$ and $AN(2)_\bn$ ($\RR^3$ when $\bn=0$) 
take the form
\bea
\{\tilde p^a,\tilde p^b\}=\tilde p_3( n^a \tilde p^b-\tilde p^a n^b) \quad \mbox{and} \quad
\{\tilde q^a,\tilde q^b\}=\tilde q_3\epsilon^{abc}\tilde q_c,
\eea 
with $\tilde p_3$ and $\tilde q_3$ defined in terms of \,$\tilde p^a$ and $\tilde q^a$  as in \eqref{p3q3def}.
The symplectic leaves of the Poisson manifold
$AN(2)_\bn$ ($\RR^3$ when $\bn =0$) are the orbits 
under the dressing action \eqref{dressr} and
the symplectic leaves of the Poisson manifold $\HH_1 $ are the
orbits  of the dressing action \eqref{dressv} for $\Lambda \neq 0$
and \eqref{dressvzero} for $\Lambda =0$.
\end{theorem}

{\bf Proof}: This is an application of standard results in the 
theory of Poisson-Lie groups. The formula for the Poisson brackets 
follows from the fact that, as a  Poisson manifold,
 a classical double  equipped with the Sklyanin bracket 
is a direct product of the factor groups equipped with their respective Poisson
structures; see e.g. Proposition 8.4.5 in \cite{Majid}. The characterisation
of symplectic leaves as orbits of dressing transformations goes back to
\cite{setishan}, see \cite{K-S} for a pedagogical exposition and further 
references. \hfill $\Box$

The other Poisson structure relevant to the description of the phase
space of 3d gravity is the Heisenberg double Poisson structure
introduced by Semenov-Tian-Shansky \cite{setishan} which is related
to the contributions of the handles. The Heisenberg double $D_+(G)$
of a Poisson-Lie group $G$ is the group $G\times G$ equipped with
the unique Poisson structure such that the canonical embeddings
$G\rightarrow G\times G$ and $G^*\rightarrow G\times G$ of the
Poisson-Lie group and its dual are Poisson maps. The former is
simply the diagonal map $G\ni g\mapsto (g,g)\in G\times G$, the
latter is defined in terms of the factorisation of $G$ into the two
subgroups associated to its classical $r$-matrix. Denoting these
subgroups by $H$ and $K$, one has $G^*=H\times K$ as a group, and
the embedding of $G^*$ into $G\times G$ is given by $H\times
K\ni(h,k)\mapsto(h,k^\inv)$. It is shown in \cite{setishan} that the
Heisenberg double Poisson structure $D_+(G)$ is given by the Poisson
bivector
\begin{align}
\label{hdbivect} B_{D_+}=&\tfrac{1}{2}r^{ba}\left( X_a^{1R}\wedge
X_b^{1R}+X_a^{1L}\wedge X_b^{1L}\right)+\tfrac{1}{2}r^{ba}\left(
X_a^{2R}\wedge X_b^{2R}+X_a^{2L}\wedge
X_b^{2L}\right)\\
+&r^{ba}\left( X_a^{1R}\wedge X_b^{2R}+X_a^{1L}\wedge
X_b^{2L}\right),\nonumber
\end{align}
where $X_a$, $a=1,\ldots,\text{dim}(\gothg)$, is a basis of
$\gothg=\text{Lie}(G)$, $X_a^{iR}$, $X_a^{iL}$ denote the associated
left- and right-invariant vector fields on the two components of the
group and $r=r^{ab} X_a\oo X_b$ its classical $r$-matrix.

For the gravity Lie groups $G=\HH_1(R_\Lambda)$, we have $H=\HH_1$
and $K=AN(2)$ (or $K=\RR^3$ when $\bn=0$), and the canonical embeddings of $\HH_1(R_\Lambda)$ and
$\HH_1(R_\Lambda)^*=\HH_1\times AN(2)$ (or $\HH_1(R_\Lambda)^*=\HH_1\times \RR^3$) are given by
\begin{align}
&\HH_1(R_\Lambda)\rightarrow \HH_1(R_\Lambda)\times \HH_1(R_\Lambda),
& &g\mapsto(g,g),\\
&\HH_1(R_\Lambda)^*\rightarrow
\HH_1(R_\Lambda)\times \HH_1(R_\Lambda), & &(u,s)\mapsto(u,s^\inv).
\end{align}
Applying Semenov-Tian-Shansky's Poisson bivector \eqref{hdbivect} to
the coordinate functions defined above, we obtain the following
theorem characterising the Heisenberg double of the Poisson-Lie
groups $\HH_1(R_\Lambda)$.

\begin{theorem}
\label{hdth}

Writing elements of $\HH_1(R_\Lambda)\times \HH_1(R_\Lambda)$
as $(g_1,g_2)$, with $g_1,g_2\in \HH_1(R_\Lambda)$, and using the
coordinate functions
\begin{align}
&p^a((g_1,g_2))=-2\Pi(u(g_1)\cdot e^a), &  &k_a((g_1,g_2))=-2\Pi(u(g_2)\cdot e^a),\\
&q^a((g_1,g_2))=-2\text{Im}_\tl \Pi(s(g_1)\cdot e^a), &
&l_a((g_1,g_2))=-2\text{Im}_\tl\Pi(s(g_2)\cdot e^a),
\end{align}
with $u, s$ defined as in \eqref{factform1}, \eqref{factform3}, and
the associated functions $p_3,q_3,k_3,l_3$ defined as in
\eqref{p3q3def},  the Heisenberg double Poisson
structure associated to the classical $r$-matrix \eqref{poincr}
takes the form
\begin{align}
&\{p^a,p^b\}_{D_+}=p_3\left( n^a p^b-p^a n^b\right),\qquad\{q^a,q^b\}_{D_+}=-q_3\;\epsilon^{abc}q_c,\\
&\{p^a,q^b\}_{D_+}=p_3q_3\,\eta^{ab}+\tfrac{1}{2}p_3\left(n^aq^b-\bq\bn\,\eta^{ab}\right)-\tfrac{1}{2}q_3\epsilon^{abc}p_c-\tfrac{1}{4}\epsilon^{adf}p_f\left(n^dq^b-\bq\bn\,\eta^{db}\right),\nonumber\\
\intertext{} &\{k^a,k^b\}_{D_+}=k_3\left( n^a k^b-k^a
n^b\right),\qquad\{l^a,l^b\}_{D_+}=-l_3\;\epsilon^{abc}l_c,\nonumber\\
&\{k^a,l^b\}_{D_+}=k_3l_3\,\eta^{ab}+\tfrac{1}{2}k_3\left(n^al^b-\bl\bn\,\eta^{ab}\right)-\tfrac{1}{2}l_3\epsilon^{abc}k_c-\tfrac{1}{4}\epsilon^{adf}k_f\left(n^dl^b-\bl\bn\,\eta^{db}\right),\nonumber\\ 
\intertext{} &\{q^a,k^b\}_{D_+}=0,\nonumber\\
&\{p^a,k^b\}_{D_+}=\tfrac{1}{2}\epsilon^{acd}p_ck_d\left(\tfrac{1}{4}\bk\bn
k^b-n^b\right)+\tfrac{1}{2}p_3^2
k^b\epsilon^{acd}p_cn_d-\tfrac{1}{2}p_3k_3 k^b\epsilon^{acd}k_cn_d+\tfrac{1}{4}k^ak^b(k_3\bp\bn-p_3\bk\bn)\nonumber\\
&\qquad\qquad-n^ak^b(p_3^3+\tfrac{1}{4}k_3\bp\bn)+p_3k^an^b,\nonumber\\
&\{q^a,l^b\}_{D_+}=\frac{l_3+\tfrac{1}{2}\bl\bn}{q_3+\tfrac{1}{2}\bq\bn}(\tfrac{1}{2}q^b\epsilon^{acd}q_cn_d-q_3\epsilon^{abc}q_c)-l^b\epsilon^{acd}q_cn_d,\nonumber\\
&\{p^a,l^b\}_{D_+}=\frac{l_3+\tfrac{1}{2}\bl\bn}{q_3+\tfrac{1}{2}\bq\bn}(p_3n^aq^b+(q_3-\tfrac{1}{2}\bq\bn)(p_3\eta^{ab}-\tfrac{1}{2}\epsilon^{abc}p_c)+\tfrac{1}{2}q^b\epsilon^{acd}p_cn_d)-\frac{q_3p_3
n^al^b}{q_3+\tfrac{1}{2}\bq\bn}\nonumber\\
&+(l_3-\tfrac{1}{2}\bl\bn)(
(p_3(1-\tfrac{1}{2}\bk^2)+\tfrac{k_3}{2}\bp^2)\eta^{ab}
-(k_3p_3-k_3^2+\tfrac{1}{2})\epsilon^{abc}p_c+\tfrac{1}{2}p_3k^ak^b-\tfrac{1}{2}k_3p^ap^b-\tfrac{1}{4}k^b\epsilon^{acd}p_ck_d)\nonumber\\
&+(p_3k_3^2-\tfrac{p_3}{2})n^al^b+\tfrac{p_3}{2}\bk\bn
k^al^b+k_3(p_3-k_3)l^b\epsilon^{acd}p_cn_d-\tfrac{1}{8}\bk\bn
l^b\epsilon^{acd}p_ck_d.\nonumber
\end{align}
\end{theorem}

{\bf Proof:} \; The formulas in the theorem can be  checked  by straightforward but lengthy computation. One
inserts the
 classical $r$-matrix \eqref{poincr} into \eqref{hdbivect}
 and applies the resulting expression to the coordinate
functions $p^a,q^a, k^a,l^a$ on the two copies of
$\HH_1(R_\Lambda)$. This yields
\begin{align}
&\{ p^a, p^b\}_{D_+}=\tfrac{1}{2}\left( J_c^L  p^a S^c_L p^b - J_c^L
p^b S^c_L p^a\right)=\{ p^a, p^b\}_S=-\{ p^a,
p^b\}_D,\\
&\{ q^a, q^b\}_{D_+}=\tfrac{1}{2}\left( J_c^R  q^a S^c_R q^b - J_c^R
q^b S^c_R q^a\right)=-\{ q^a, q^b\}_S=-\{ q^a,
q^b\}_D,\\
&\{ p^a, q^b\}_{D_+}=\tfrac{1}{2}\left( J_c^R  p^a S^c_R q^b + J_c^L
p^a S^c_L q^b\right),
\\ &\{ p^a, k^b\}_{D_+}= J_c^L p^a
S^c_L k^b,\qquad \{ p^a, l^b\}_{D_+}= J_c^L p^a S^c_L l^b+ J_c^R p^a
S^c_R
l^b,\\
&\{ q^a, l^b\}_{D_+}= J_c^R q^a S^c_R l^b,\qquad\;\; \{ q^a,
k^b\}_{D_+}=0.
\end{align}
To evaluate these expressions further one makes use of the
 formulas \eqref{lorvecll}, \eqref{transvecrr},
\eqref{jrp}, \eqref{jrq}, \eqref{slp}, \eqref{slq}  for the action
of the left- and right-invariant vector fields on $\HH_1(R_\Lambda)$
 on the coordinate functions .\hfill $\Box$

Theorem \ref{dualskylth} and Theorem \ref{hdth} provide explicit
expressions for the Sklyanin Poisson-Lie structure, the dual
Poisson-Lie structure and the Heisenberg double Poisson structure
associated to the local isometry groups arising in 3d gravity for
Lorentzian signature and the Euclidean case with $\Lambda\leq 0$.

From the viewpoint of 3d gravity this amounts to a complete
parametrisation of phase space and Poisson structure for spacetimes
of arbitrary genus and with an arbitrary number of massive spinning
particles. Not only are the resulting expressions for the Poisson
structure of a rather simple form, but the structural similarities
of the theory for different signatures and signs of the cosmological
constant are readily apparent. The resulting expressions for the
Poisson structure take the same form and the dependence on the
cosmological constant is explicit and encoded in the vector $\bn$
satisfying $\bn^2=-\Lambda$. The description thus unifies the
description of phase space and Poisson structure for different
signatures and signs of the cosmological constant and allows one to
investigate all cases in a common framework in which the cosmological
constant appears as a parameter.

\section*{Acknowledgements}
Research at Perimeter Institute for Theoretical Physics is supported
in part by the Government of Canada through NSERC and by the
Province of Ontario through MRI.

\end{document}